\documentclass[bibyear]{aa}

\usepackage{graphicx}
\usepackage{txfonts}
\usepackage[breaklinks=true]{hyperref}
\hypersetup{
    colorlinks = true,
    citecolor = black,
    linkcolor=blue,
    urlcolor = blue
    }

\usepackage{natbib, twoopt}
\usepackage{textcomp, gensymb}
\usepackage{multirow}
\usepackage{longtable}
\usepackage{supertabular}
\usepackage{amsmath}
\usepackage{amssymb}
\usepackage[T1]{fontenc}

\usepackage{booktabs}
\usepackage{tabularx}
\usepackage{array}

\begin{document}
    \title{Asteroseismic Analysis of the Merger Product Red Giant in the $\gamma$~Persei System}
    \titlerunning{Asteroseismic Analysis of $\gamma$~Per~A}

   \author{Roz\'{a}lia Z. \'{A}d\'{a}m \inst{1}\fnmsep\inst{2}
          \and
          L\'{a}szl\'{o} Moln\'{a}r\inst{1}\fnmsep\inst{2}\fnmsep\inst{3}
          \and
          R\'{o}bert Szab\'{o}\inst{1}\fnmsep\inst{2}\fnmsep\inst{3}
          \and
          Csilla Kalup\inst{1}\fnmsep\inst{2}
          \and
          Frank Grundahl\inst{4}
          \and
          Daniel Huber\inst{5}
          \and
          Mads Skakke Fredslund\inst{4}
          \and
          Pere L. Pallé\inst{6}\fnmsep\inst{7}
          \and
          D\'ora Tarczay-Neh\'ez\inst{1}\fnmsep\inst{2}
          \and
          Jonatan Rudrasingam\inst{4, 8, 9}
          }

    \institute{Konkoly Observatory, Research Centre for Astronomy and Earth Sciences, HUN-REN, MTA Centre of Excellence, Konkoly Thege Mikl\'os \'ut 15-17, H-1121 Budapest, Hungary\\
              \email{adam.rozalia@csfk.org}
        \and
            ELTE E\"otv\"os Lor\'and University, Institute of Physics and Astronomy, H-1117 P\'azm\'any P\'eter s\'et\'any 1/A, Budapest, Hungary
        \and
            MTA–HUN-REN CSFK Lendület ``Momentum'' Stellar Pulsation Research Group
        \and
            Stellar Astrophysics Centre (SAC), Department of Physics and Astronomy, Aarhus University, Ny Munkegade 120, DK-8000 Aarhus, Denmark
        \and
            Institute for Astronomy, University of Hawai'i, 2680 Woodlawn Drive, Honolulu, HI 96822, USA
        \and
            Instituto de Astrofísica de Canarias, 38200 La Laguna, Tenerife, Spain
        \and
            Universidad de La Laguna (ULL), Departamento de Astrofísica, 38206 La Laguna, Tenerife, Spain
        \and
            Sydney Institute for Astronomy, School of Physics, University of Sydney, Sydney, NSW 2006, Australia
        \and
            School of Mathematical and Physical Sciences, Macquarie University, 12 Wally's Walk, Macquarie Park, NSW 2113, Australia
            }

    \date{Received \today; accepted YYY}
    
    \abstract
    {$\gamma$ Persei is a long-period eclipsing binary system ($P\approx 14.6$ years) containing a red giant primary, and it is well known for its multi-faceted classification as a visual and spectroscopic binary. Its brightness and binary nature together make it a valuable target for both photometric and spectroscopic studies, particularly in the context of asteroseismology and stellar evolution, as the primary star likely formed through a stellar merger.}
    {We aim to determine the seismic parameters $\nu_{\rm max}$, $\Delta \nu$, and the oscillation amplitudes of the primary component, an evolved giant, to estimate its seismic mass -- which we can compare to its estimated dynamic mass.}
    {We use Transiting Exoplanet Survey Satellite (TESS) data obtained during Sectors 58, 85, and 86 and to complement the space-based observations, we incorporate high-resolution RV measurements acquired by the Stellar Observations Network Group (SONG) during two distinct epochs; 2017 and 2024.}
    {We successfully detect solar-like oscillations in $\gamma$~Per and infer a seismic mass of $3.25\pm0.13$ M$_\odot$, which is slightly below the dynamical mass. We find the photometric oscillation amplitudes to be significantly lower than predicted from scaling relations, but in line with other high-mass red giants. We also find that radial velocity amplitudes along the Hertzsprung-Russell diagram cannot be fitted uniformly with current scaling relations.}
    {}

   \maketitle

\section{Introduction}
$\gamma$ Persei is an eclipsing binary system with an almost 15 year-long orbital period, which has been studied from many distinct aspects during the last two century \citep{Campbell_1909, Pourbaix_1999, Pourbaix_2000, Hartkopf_2001, Pourbaix_2004, Picotti_2020, Diamant_2023}.
The system is bright enough to be visible to the naked eye with $V=2.93$\,mag. Yet, only two primary eclipses were known in the literature for some time, from 1990 by \cite{Griffin_1994} and 2019 by \cite{Diamant_2023}. Recently, \cite{Rozi-2025} extended this picture by presenting the main eclipse from 2005, and the secondary eclipse from 2006.

The first photometric analysis of $\gamma$ Per was done by \cite{Griffin_1994}, where they published light curves in Johnson $U$, $B$ and $V$ passbands, and modelled the eclipse geometry. From that they offered radii for both components of the system ($22.2 \ {\rm R}_\odot$ and $3.9 \ {\rm R}_\odot$). The masses of the stars in the system were calculated by \citet{McLaughlin1948} and \citet{McAlister-1982} to be $M_1 = 4.73$\,M$_\odot$ and $M_1 = 2.75$\,M$_\odot$. Later, \citet{Popper-1987} derived significantly lower values of $M_1 = 3.0$\,M$_\odot$ and $M_1 = 2.0$\,M$_\odot$. While analysing the chromosphere, \cite{Diamant_2023} determined new and more precise astrophysical parameters for the system, such as luminosity, effective temperature, surface gravity, radius, and mass for primary and secondary, the latter values falling to $3.6 \pm 0.2$\,M$_\odot$ and $2.4 \pm 0.2$\,M$_\odot$.

An interesting aspect of the system was raised by \citet{Griffin-2007} and later by \citet{Diamant_2023}, who found that the components of the system cannot be fitted with a single isochrone. The much younger age of the primary component suggests that the red giant went through a fundamental alteration in its structure and evolution at some point. We investigate a possible merger scenario of two main sequence (MS) stars to form the current primary based on fits to isochrones and stellar tracks in a companion paper \citep{TND-2026}. In that paper we show that the true age of the secondary, and thus the system, is $\sim 750-900$\,Myr, with the merger creating the current primary star occurring at age $\sim 500-775$\,Myr. Since this merger happened between two MS stars, this rejuvenated new primary is now largely indistinguishable from a normal red giant branch (RGB) star, but with an apparent single-star age corresponding to $\sim 280-350$\,Myr.

Since the primary star of the $\gamma$~Per system is a red giant, it is expected to show solar-like oscillations. The Transiting Exoplanet Survey Satellite \citep[TESS,][]{Ricker2015} has been collecting high-precision, continuous photometry from the whole sky, and detected solar-like oscillations in numerous stars across the Hertzsprung--Russell diagram (HRD), permitting population-level asteroseismic studies \citep[e.g.,][]{Hon-2021,Hon-2022,Hatt-2023,Zhou-2024,Grusnis-2025,Lund-2025}. These stars also include very bright targets, which can be followed up by spectroscopic observations, and their oscillations can be studied in the radial velocity (RV) variations as well, allowing for detailed stellar characterization \citep[e.g.,][]{Gupta_2022,Knudstrup-2023,Kjeldsen-2025,Tang-2026}. Though, $\gamma$~Per was not part of any prior study yet.

In this paper, we aim to obtain a better picture of $\gamma$ Per by offering independent mass estimates for its red giant primary component through asteroseismic analysis of high-precision photometric and RV observations.
In Section~\ref{sec:data}, first, we present the measurements and facilities we used, then we describe the varied aspects of our analysis. In Section~\ref{sec:results} we detail our results and compare $\gamma$~Per to other oscillating stars with seismic RV data, while in Section~\ref{sec:discussion} we aim to summarize and highlight the importance of our findings.

\section{Data Processing} \label{sec:data}
Continuous, high-precision photometric data was collected by the TESS mission during multiple sectors. RV observations were collected by the Stellar Observations Network Group (SONG) project \citep{Grundahl-2008}.  

\subsection{TESS Observations}
TESS observes a $24^\circ \times 96^\circ$ degree area of the sky with four 10 cm telescopes nearly continuously for one lunar orbit (27 days). During this time TESS orbits Earth twice, with data download periods during perigee. The field-of-view rotated around the sky with one camera fixed on the ecliptic poles for one year, flipping between the Southern and Northern hemispheres after one year in the early phase in the mission, with different orientations also implemented later on. This observing strategy leads to multiple-sector observations of certain targets where camera pointings can overlap, with long data gaps between repeated visits to the same areas. We processed four visits to $\gamma$~Per in this work.

TESS observed $\gamma$~Per during Sector 18 in 2019, which serendipitously coincided with a primary eclipse \citep{Rozi-2025}. Sector 18 is also the shortest piece of continuous light curve TESS has collected for $\gamma$~Per: $24.4$ days of data, with a 2.5-day gap in the middle. Since during this phase the red giant branch (RGB) component is eclipsing the hotter but smaller secondary star, oscillations are visible unimpeded during the eclipse. We attempted to determine global oscillation parameters from S18, but found the results to be inconclusive.

The space telescope observed the system again during Sector 58 in late 2022, collecting $27.7$ days of almost completely continuous data. Since solar-like oscillations are stochastic by nature, short segments may not be representative of the overall seismic profile of the star. Therefore, it is important that we have access to a sector distinct from the S85--86 data, and that we can compare the PSD of the star from different points of time. The PSD of S58 is shown in Figure \ref{fig:tess_s58_psd}. Signs of a power excess are visible between 12--28\,$\micro$Hz, which is in agreement with our results based in S85--86 in Section~\ref{sec:mass}, but we could not determine a reliable $\nu_{\rm max}$ value from this sector alone.

\begin{figure*}
    \includegraphics[width=\linewidth]{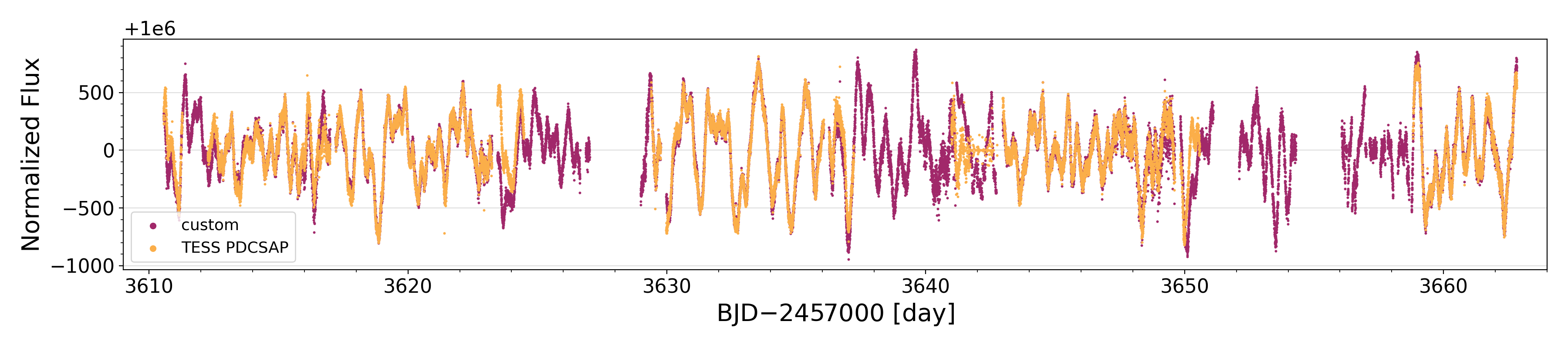}
    \includegraphics[width=\linewidth]{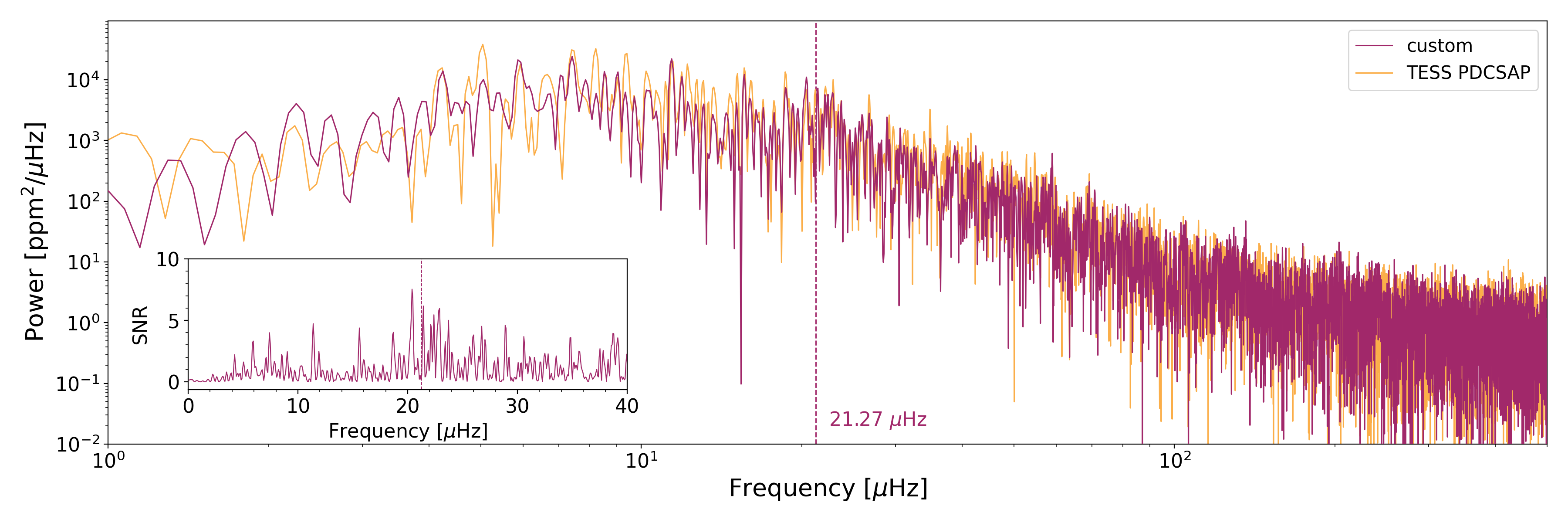}
    \caption{We produced two light curves from data points collected by TESS in Sectors 85 and 86. \textit{Top:} our custom light curve with magenta, and the downloadable PDCSAP light curve with orange. \textit{Bottom:} the power spectral density (PSD) of these light curves with the respective colours, and with an inset of the SNR of the PSD of the custom light curve.}
    \label{fig:s85-86}
\end{figure*}

Finally, $\gamma$~Per was observed again in Sectors 85 and 86 in late 2024. Sectors 85--86 are the most important for asteroseismology since they provide us with a quasi-continuous light curve that spans 53 days. TESS data are available in multiple forms, including light curves from various pipelines. The processed PDCSAP (pre-search data conditioned simple aperture photometry) light curve released by the TESS Science Processing Operations Center (SPOC) contains multiple gaps in the TESS light curve, where cadences failed various quality cuts, limiting the duty cycle to 61\%. In order to limit these gaps, we downloaded the full SAP (simple aperture photometry) light curve that includes every photometric data point. We then manually removed segments most affected by scattered light and large jumps in intensity. The quality of S86 data decreased towards the ends of the orbit: we smoothed the end of the first orbit of S86 to recover the relatively slow variations caused by the oscillations. This increased our temporal coverage significantly, to a duty cycle of 85\%. We refer to this latter light curve as our custom light curve throughout the paper. The differences between the two light curves and their PSDs are shown in Fig.~\ref{fig:s85-86}.

\subsection{SONG Project}
The SONG project aims to study the internal structure and evolution of stars using asteroseismology, such as solar-type oscillations, and to search for and characterize Earth-mass exoplanets. It was launched in 2006 by astronomers at Aarhus University and the University of Copenhagen, and is an international collaborative project.
The idea behind the project was to create a network of small telescopes, consisting of eight relatively inexpensive, ultra-modern robotic telescopes.

We used the 1m Hertzsprung SONG telescope, the prototype node of the SONG network installed in 2014, and the SONG spectrograph at Observatorio del Teide on Tenerife (see \citet{Grundahl2017} and \citet{Andersen2016} for general information on SONG) to obtain precise radial velocities during two observing periods, in 2017 and 2024.
The 2017 observations were targeting $\gamma$~Per as an asteroseismic target for a possible stellar merger product (PI: Huber). The 2024 observations were proposed to coincide with the TESS observations, in order to be able to compare the RV and photometric asteroseismic signals directly (PI: Moln\'ar).

For the observations prior to 2024, we used a slit (no.~5) resulting in $R \approx 77,\!000$, and an ANDOR IKON-L detector with $2048\times 2048$ pixels. The exposure time was kept at $120$\,s. In January 2024, this camera was changed to a QHY-600Pro camera. All data from 2024 was obtained at the maximum spectral resolution of $130,\!000$ and an exposure time of $120$\,s. In order to do precise wavelength calibration, all data was obtained using an iodine cell and reduced using the code \texttt{pyodine} \citep{Pyodine2023}.  The typical radial velocity precision obtained was in the $3-4$ m/s range.

SONG data are made openly available to the SONG community through the SONG Data Archive (SODA\footnote{\url{https://soda.phys.au.dk/}}) webpage. SONG provides three data products: raw uncalibrated 2d spectra, calibrated and extracted spectra and radial velocity time series. Within a few days of the observations raw and extracted spectra are made available on SODA. Radial velocity time series on specific targets can be requested by e-mail to the Instrument Scientist.

\subsubsection{2017 and 2024 RV Data} \label{sec:song_data}
The 2017 observations consist of three segments covering 128 days, with short gaps between the segments. The data show a slow trend in the RVs, which we fitted and removed with a second-order polynomial. The first 20 days of the campaign were observed in high cadence mode with up to 250 measurements per night. Data taken after that are typically on the order of a dozen observations per night.

Somewhat surprisingly, the first segment, including the high-cadence observations, displays a strong beating pattern instead of the usual, semi-regular variations around typical periodicities and amplitudes. This beating then subsides, and the low-cadence part shows far less amplitude variations. 

\begin{figure}
    \includegraphics[width=\linewidth]{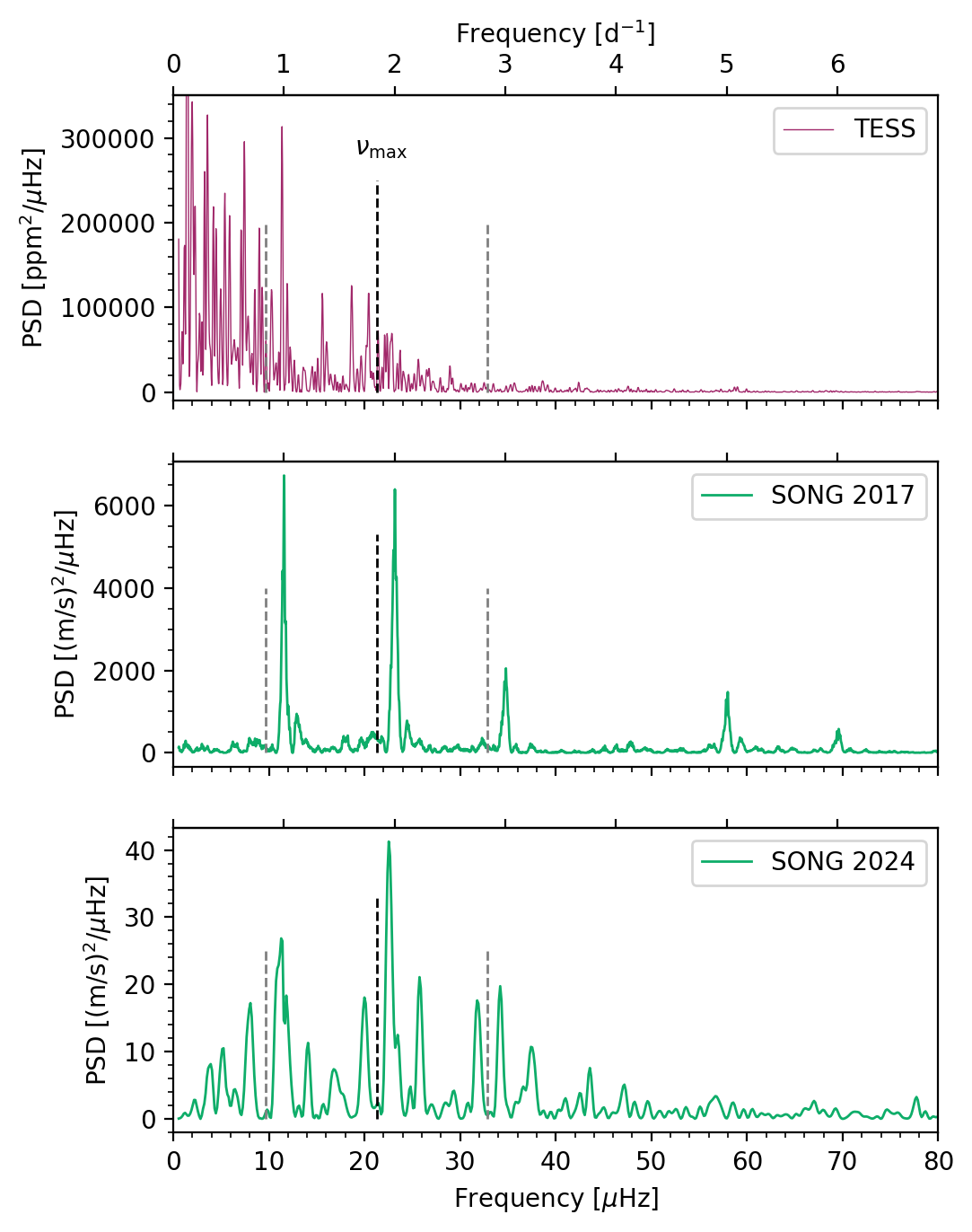}
    \caption{From top to bottom: power spectra of the TESS S85--86 custom light curve, and the 2017 and 2024 SONG measurements. Dashed black line marks the $\nu_{\rm max}$ frequency determined from the TESS data. Grey lines mark the $\pm 1.0$\,d$^{-1}$ alias values of $\nu_{\rm max}$.}
    \label{fig:tess-song_psd_comparison}
\end{figure}

This peculiarity of the early 2017 data causes the frequency spectrum to show strong, narrow signals close to integer frequencies 1.0, 2.0, 3.0\dots d$^{-1}$, instead of a broader power excess typical of solar-like oscillations, and its daily aliases (Fig.~\ref{fig:tess-song_psd_comparison}). This makes the determination of the true frequency content especially difficult. Even with observations spanning full nights, the RV curve can be folded adequately either with the 1.0 or 2.0 d$^{-1}$ frequency due to the limited phase coverage. 

We note that the fact that peaks in the frequency spectrum coincide with the alias peaks in the spectral window of the data does not mean that signals are not real, or that they are only caused by the sampling of the data, although such claims have been made in the literature occasionally. Mathematically, the discrete Fourier transform, $F_N(\nu)$, expressed as
\begin{equation}
    F_N(\nu) = \sum_{i=1}^{N} x(t_i) \ e^{2\pi\, i \nu t}   
\end{equation}
is the convolution of the window function ($W_N(\nu)$, the discrete Fourier transform of the time sampling):
\begin{equation}
    W_N(\nu) = \sum_{i=1}^{N} \ e^{2\pi\, i \nu t}
\end{equation}
and the Fourier transform of the actual, continuous physical signals in the data, \textit{F},
\begin{equation}
    F = \int_{0}^{T} x(t) \ e^{2\pi\, i \nu t} \mathrm{d}t
\end{equation}
as:
\begin{equation}
    F_N(\nu)/N = (F\ast W_N)(\nu),  
\end{equation}
where $x(t)$ and $x(t_i)$ are the continuous signal and its observations at discrete $t_i$ times. Therefore, for any significant peak to appear in the frequency spectrum, a detectable signal is required so that $F\neq0$ \citep{A-JCD-K-2010a}. Here, we are merely left with an ambiguous spectrum with confusion between the true astrophysical signal close to either 1.0 or 2.0 d$^{-1}$, and its aliases. We show the solution to this puzzle, and a possible explanation for the lack of a wider power excess, in Sect.~\ref{sec:song_alias}. 

The 2024 observing campaign was designed to be contemporaneous with the TESS S85--86 measurements. Observations ran from 2024-11-26 to 2024-12-16, but the first week was mostly lost to bad weather. Useful data was collected from day 5, resulting in a 17~d long data set, coinciding with S86. Long-term RV changes were once again removed with a second-order polynomial fit. As Fig.~\ref{fig:tess-song_psd_comparison} shows, the power spectrum of this data set shows multiple peaks repeating around 1.0, 2.0 and 3.0 d$^{-1}$. This is more reminiscent of a solar-like oscillator, but still includes the aliasing uncertainties.

\begin{figure*}
    \includegraphics[width=\linewidth]{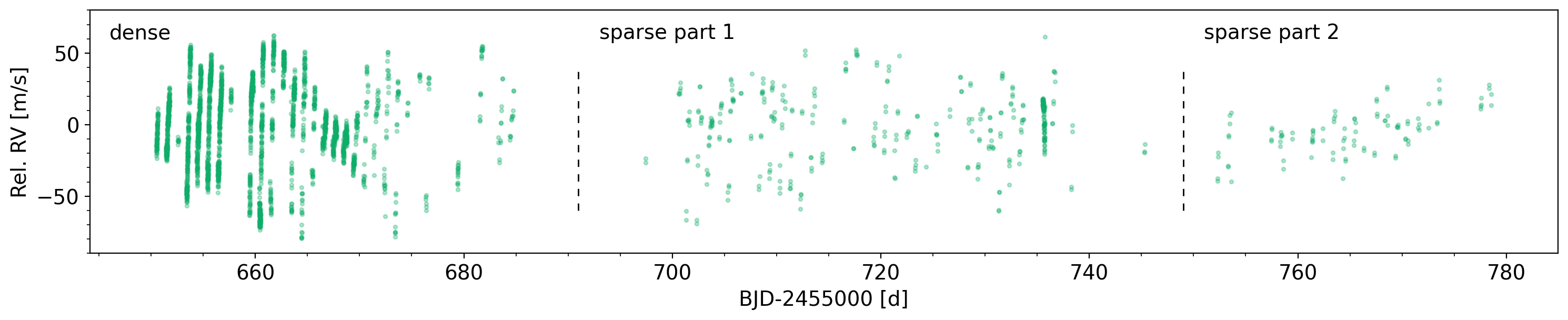}
    \caption{SONG RV measurements of $\gamma$~Per during the 2017 campaign, with dashed lines marking the separation between the dense, and the middle and late sparse data segments. }
    \label{fig:song_2017_all}
\end{figure*}

\subsubsection{Gaussian Processes Interpolation of the RV Data} \label{sec:song_gp}
We fitted the RV data with Gaussian Processes (GPs) to create continuous representations that approximate the behaviour of the signal. We used the \texttt{tinygp} python package \citep{tinygp-2024} for this purpose. GPs are stochastic models consisting of a mean function, $m(x)$ (our initial guess), and a covariance, or `kernel' function, $k(x, x')$ (the prior), and assume that the value of each data point to be a normally distributed random variable. The joint distribution then is:
\begin{equation}
    f(x) \sim \mathcal{GP}\left(m(x), k(x, x')\right).
\end{equation}
The log-likelihood function of a model fitting the data is in effect a generalization of the classical $\chi^2$ value, and our goal is to maximize this log-likelihood function.

Here, we decided to use a simple harmonic oscillator (SHO) kernel to fit the data, defined by \citep{DFM-2017} as
\begin{equation}
    \left[\frac{\mathrm{d}^2}{\mathrm{d} t^2} + \frac{\omega_0}{Q}\,\frac{\mathrm{d}}{\mathrm{d} t}
    + \omega_0^2\right]\, y(t) = \epsilon(t),
\end{equation}
where $\omega_0$ is the frequency of the undamped oscillator, $Q$ is the
quality factor of the oscillator, and $\epsilon(t)$ is a stochastic driving
force. The value of $Q$ determines the level of damping of the oscillator. For $Q<1/2$ the oscillator is overdamped, for large $Q$ values it is underdamped. At $Q=1/2$ it approaches the functional form of granulation, which can also be described by Harvey-like functions, defined by \citet{Harvey-1985}. Large $Q$ values are good models for solar-like oscillations, at which the kernel function can be described as:
\begin{eqnarray}
k(\tau) \approx
    S_0\,\omega_0\,Q\, e^{-\frac{\omega_0\,\tau}{2\,Q}}\, \cos\left(\omega_0\,\tau\right).
\end{eqnarray}

For further forms of this kernel at lower $Q$ values we refer the reader to \citet{DFM-2017}. The \texttt{tinygp} code also includes a $\sigma$ scaling parameter in the $k(\tau)$ function, that could be used for slight performance boost, as per the description of \texttt{tinygp}: we left this value at $1.0$. We set the initial frequency to $\omega_0 = 2\pi\, 2.0$\,d$^{-1}$, based on the strongest frequency of $2.0\,d^{-1}$ we found in the frequency spectrum, which is also close to the $\nu_{\rm max}$ of the photometric data set, as we show in Section \ref{sec:results}.

As the cadence of the measurements and the shape of the RV curve both changed considerably during the 2017 measurements, we split the data into three segments and fitted them separately. The individual segments are short relative to the expected lifetime of the modes. Following \citet{DFM-2017}, we found that underdamped oscillators with high $Q$ quality factors gave us the best results. For the 2024 data, we chose $Q=10$, and for the 2017 data segments we tried multiple values. For the first, better-sampled segment we show both $Q=9$ and 10 to produce two alternate fits in Fig.~\ref{fig:song_2017_dense}. For the sparser segments we tested two fits again, this time with both $Q=18$ and $20$, because lower $Q$ values tended to converge to solutions around $\sim 1.0\,d^{-1}$ dominant frequency. For the middle we show the $Q=18$ and $20$ fits, and for the last segment a $Q=20$ fit in Figs.~\ref{fig:song_2017_sparse}.

We also included a jitter term to account for the white noise present in the data by adding an extra term the diagonal elements of the covariance matrix. We extracted the maximum likelihood model and created an interpolated curve for the time span of each segment. After testing for multiple values for the jitter term between $0$ and $1.0$, we chose $\log\sigma_{\rm jitter} = 0.45$ (corresponding to $\sigma_{\rm jitter} = 1.6$\,(m/s)$^2$), based on the best visual match between the RV data and the GP-interpolated RV curve. We then compared the PSD of these GP-based curves to the PSD of the observed RV data segments.

\begin{figure*}[h!]
    \includegraphics[width=\linewidth]{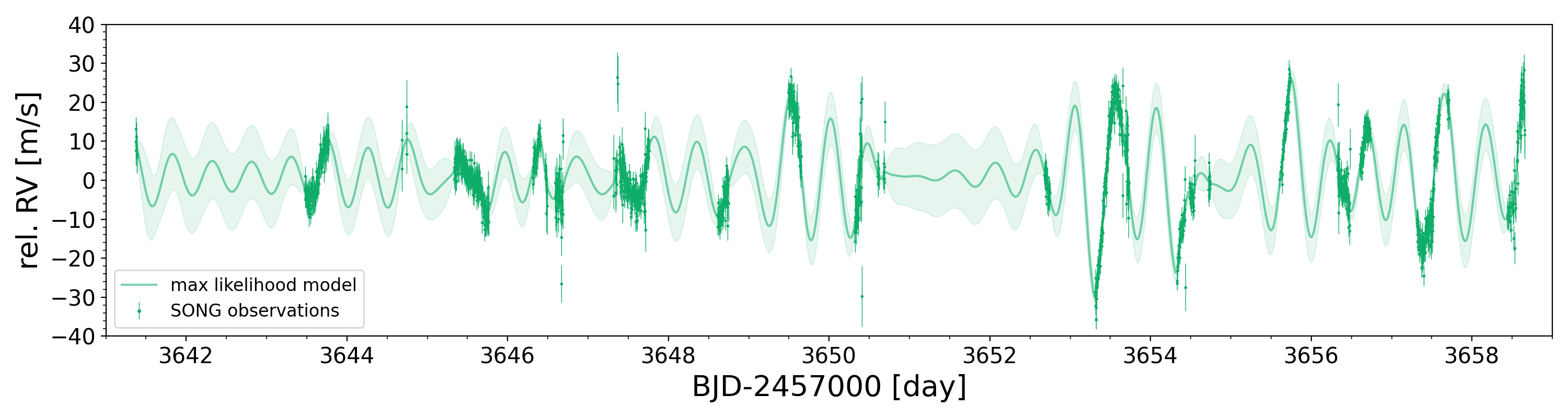}
    \includegraphics[width=\linewidth]{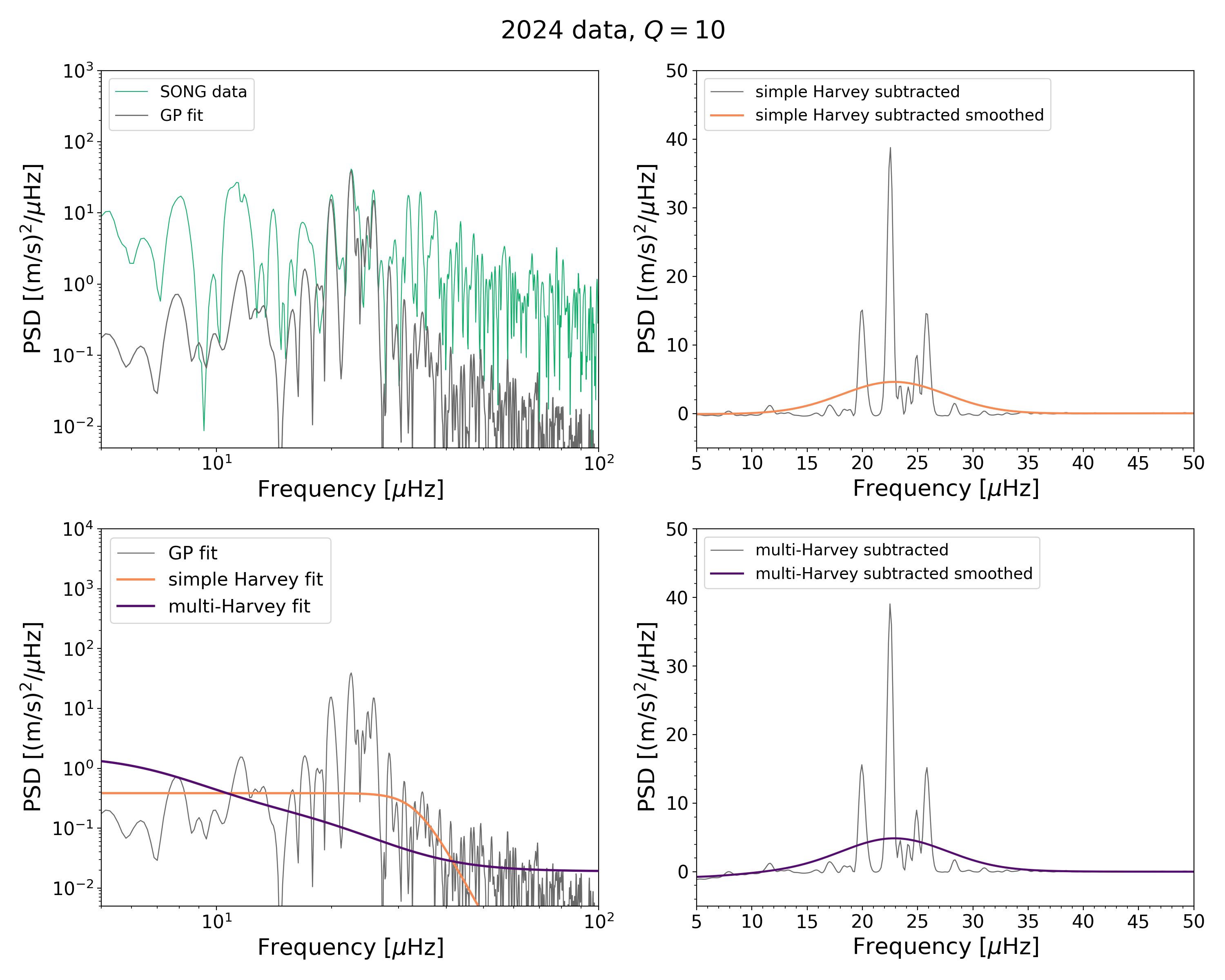}
    \caption{SONG observations from the 2024 campaign. \textit{Top panel:} we interpolated the data using the \texttt{tinygp} code whose maximum likelihood model is shown with a green line, the light green area refer to standard deviation of the model, while the observations are denoted with green dots.
    \textit{Middle left panel:} the PSD of the interpolated 2024 data (GP fit) with grey, compared to the PSD of the original observations in green. \textit{Bottom left panel:} the same PSD of the interpolated 2024 data with grey shown with the granulation background fits; a Harvey-like function with orange line, and a multi-component Harvey-like function with purple.
    \textit{Middle and bottom right panels:} we display the PSD after subtracting the simple Harvey background fit in grey in the upper plot and PSD after subtracting the multi-component Harvey in the lower one (here on a linear scale). The orange and purple lines refer to smoothed PSDs, respectively.}
    \label{fig:song_2024}
\end{figure*}

We decided to work with the time-domain representation of the model, interpolating over the gaps, in order to be able to compare the PSDs of the interpolated representation (instead of the PSD of the SHO kernel) and the PSD of the RV data more directly, and to estimate the mode-averaged oscillation amplitudes in the same way as it was done for other data sets.
Time-domain representations of the models of the 2017 data are displayed in Fig.~\ref{fig:song_2017_dense}, their respective PSDs used for amplitude calculation are shown in Figs.~\ref{fig:song_2017_densea_backfits} and~\ref{fig:song_2017_denseb_backfits}. 
The fit for the 2024 data is shown in the upper panel of Fig.~\ref{fig:song_2024}, where green dots refer to the observations, the green shaded area refer to uncertainty of the \texttt{tinygp} model, and the green line indicates the maximum likelihood model. As the figures show, both the oscillation signals were reproduced well by the GP models. At the same time the granulation background became more visible, although the model could be over-smoothing the highest frequency range.

\section{Asteroseismic Analysis}
\label{sec:seismic}
Our main aim was to characterize the solar-like oscillations in $\gamma$~Per~A, and to derive the physical parameters of the RGB component through asteroseismic analysis. This can be achieved in two ways. One is peakbagging, i.e., determining the frequencies of individual modes and comparing those to oscillation frequencies of stellar models. We found the TESS light curves to be too short for this type of analysis. The other way is to determine the global characteristics of the oscillation spectrum, and use scaling relations to infer stellar parameters. In this latter case, our aim is to determine the parameters $\nu_{\rm max}$, the frequency of maximum oscillation power, and $\Delta \nu$, the average large frequency separation between successive radial mode orders.

\subsection{Scaling Relations} \label{sec:scale}
Scaling relations for $\Delta \nu$ and $\nu_{\rm max}$ were first proposed by \cite{Ulrich_1986} and \cite{Brown_1991}, respectively, and then summarized by \citet{Kjeldsen-Bedding_1995}. They showed that asteroseismic parameters of other stars can be inferred by scaling the solar values based on relations between solar and stellar mass, radius and effective temperature (or, alternatively, luminosity) values. \cite{Kallinger_2010} and \citet{Miglio-2012} rearranged these equations to infer stellar masses, in various forms, which we completed with the correction factors:
\begin{align}
    \frac{M}{M_\odot} &= \left(\frac{f_{\nu_{\rm max}}\,\nu_{\rm max}}{\nu_{\rm max, \odot}}\right)^3 \left(\frac{\Delta \nu}{f_{\Delta \nu}\,\Delta \nu_{\odot}}\right)^{-4} \left(\frac{T}{T_\odot}\right)^{1.5} \label{eq1} \\
    \frac{M}{M_\odot} &= \left(\frac{\Delta \nu}{f_{\Delta \nu}\,\Delta \nu_{\odot}}\right)^2 \left(\frac{L}{L_\odot}\right)^{1.5} \left(\frac{T}{T_\odot}\right)^{-6} \label{eq2}\\
    \frac{M}{M_\odot} &= \left(\frac{f_{\nu_{\rm max}}\,\nu_{\rm max}}{\nu_{\rm max, \odot}}\right) \left(\frac{L}{L_\odot}\right) \left(\frac{T}{T_\odot}\right)^{-3.5} \label{eq3} \\
    \frac{M}{M_\odot} &= \left(\frac{f_{\nu_{\rm max}}\,\nu_{\rm max}}{\nu_{\rm max, \odot}}\right)^{2.4} \left(\frac{\Delta \nu}{f_{\Delta \nu}\,\Delta \nu_{\odot}}\right)^{-2.8} \left(\frac{L}{L_\odot}\right)^{0.3} \label{eq4}
\end{align}

Since $\gamma$ Per A is a RGB star, we used correction factors for both $\Delta \nu$ and $\nu_{\rm max}$. The $\Delta \nu$ correction factor is described as
\begin{equation}
    f_{\Delta \nu} = \left( \frac{\Delta \nu}{135.1 \ \micro\rm{Hz}}\right ) \left(\frac{\rho}{\rho_\odot}\right)^{-0.5}
\end{equation}
by \cite{Sharma_2016}, where $135.1 \ \micro$Hz refers to $\Delta \nu_\odot$. Using the \texttt{Asfgrid} code by \cite{Stello_Sharma_2022}, with settings \texttt{evstate=1}, \texttt{logz=0.009}, \texttt{teff=4970}, \texttt{mass=3.6}, and \texttt{logg=2.23},  we calculate $f_{\Delta \nu}=1.01099$ for $\gamma$~Per.
The correction factor for $\nu_{\rm max}$ is presented by \cite{Pinsonneault_2025} in Eq.~(4) as
\begin{equation}
    f_{\nu_{\rm max}} = (1+p)^{-1},
\end{equation}
where 
\begin{equation}
    p = a \ (\ln \nu_{\rm max})^3 + b \ (\ln \nu_{\rm max})^2 + c \ (\ln \nu_{\rm max}) + d,
\end{equation}
whose $a$, $b$, $c$ and $d$ parameters are based on models published by \cite{Sharma_2016} and presented in Table~2 in \cite{Pinsonneault_2025}. 
Using the $\nu_{\rm max}$ and $\Delta \nu$ derived from the custom light curve, we got $f_{\nu_{\rm max}} = 1.002505$.
We note that there are more recent works on the subject, e.g., \cite{Li_2023} derived a more sophisticated formula with respect to stellar properties, but their work unfortunately does not cover the higher-mass range. We apply these scaling relations in Section \ref{sec:mass}.

\subsection{\texttt{PySYD} Analysis} \label{sec:pysyd}
We estimated initial values for $\nu_{\rm max}$ and $\Delta \nu$ from the TESS data with the built-in asteroseismology module of \texttt{lightkurve} \citep{Lightkurve_2018} first. More accurate values, however, require more sophisticated tools. Therefore, we subsequently analysed the light curves with the \texttt{pySYD} software \citep{chontos-2022}. This tool was specifically developed to determine the global asteroseismic parameters from the power spectra of solar-like oscillators. After an initial fit it masks the detected (or input) region around the power excess, and fits the granulation background with multi-component Harvey-like functions \citep{Harvey-1985}. It then subtracts the background model, and determines $\nu_{\rm max}$ from a heavily smoothed power spectrum. The $\Delta \nu$ parameter is determined from the autocorrelation function of the background-subtracted spectrum. For low SNR power excesses, \texttt{pySYD} often struggles with instabilities during the Monte Carlo sampling routine, which causes it to fail.
Furthermore, \texttt{pySYD} was primarily designed for MS stars, but for stars with low $\nu_{\rm max}$ ($ \lesssim50\,\micro$Hz) and short baselines, a Harvey-like function fitted to the PSDs does not necessarily converge. This can lead to failed runs or flawed $\nu_{\rm max}$ measurements. Therefore, \citet{Maddy2022} developed and validated a new method of measuring $\nu_{\rm max}$, which involves fitting a linear function between either side of the power excess window, but not fitting the rest of the PSD. This method provides accurate, consistent and stable results in \texttt{pySYD}.

\begin{figure}
\centering
\includegraphics[width=0.65\linewidth]{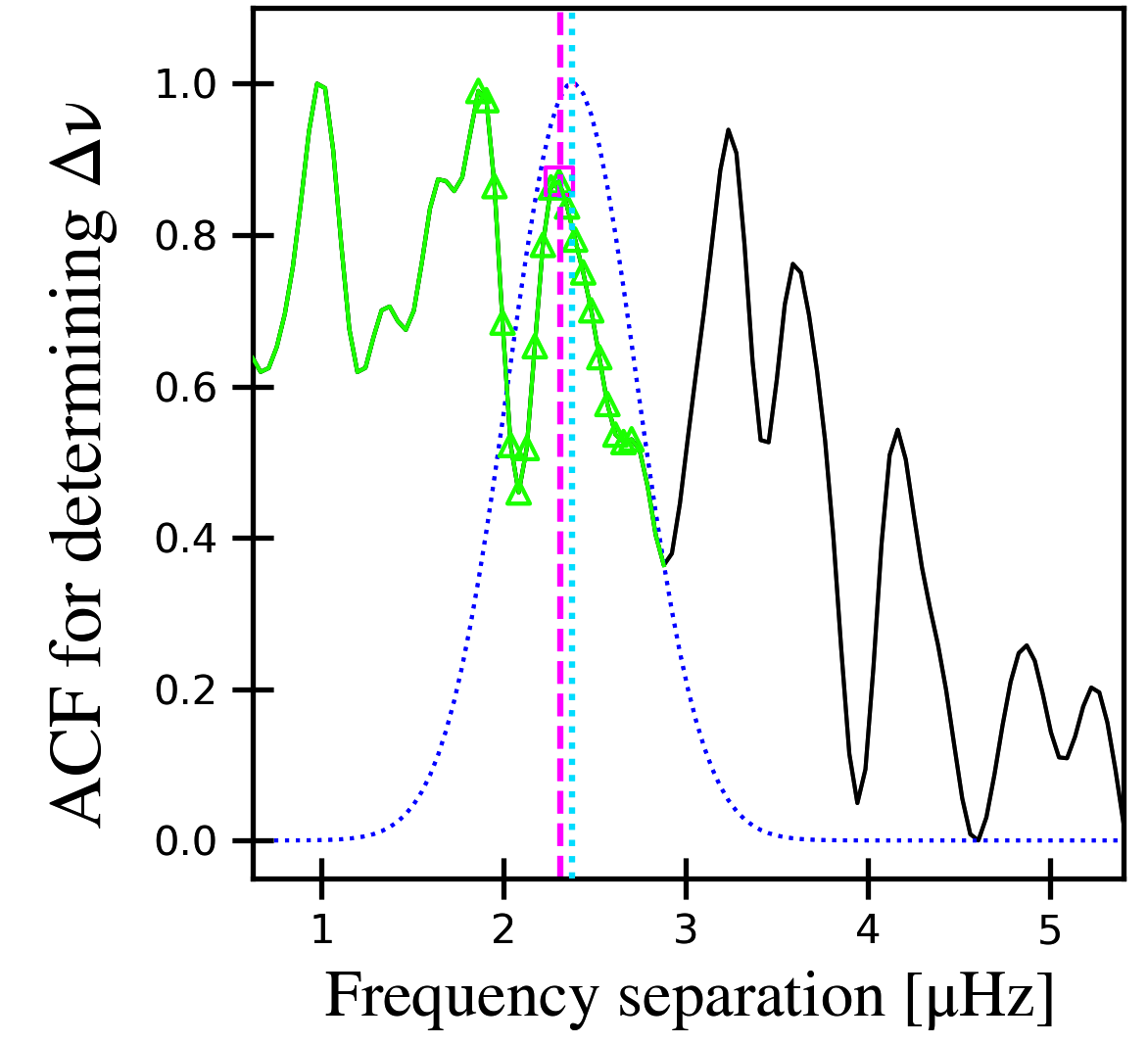}
\caption{The ACF, as calculated by \texttt{pySYD} to determine $\Delta \nu$. The blue dashed line shows the prior based on the chosen $\Delta \nu(\nu_{\rm max})$ scaling relation with the vertical cyan line marking its centre. The solid green line and symbols mark the search range within the ACF, and the purple dashed line is the identified $\Delta \nu$ value based on the peak within the prior.}
\label{fig:acf}
\end{figure}

The search range for $\Delta \nu$ is limited based on the $\nu_{\rm max}$--$\Delta \nu$ scaling relation \citep{Stello-2009}. The exact form of this scaling relation, however, is sensitive to evolutionary stage and stellar mass. Since $\gamma$~Per is known to be a high-mass red giant, we used a scaling relation of $\Delta\nu = 0.24\,{\nu_{\rm max}}^{0.75}$ to define the search range, based on the findings of \citet{Huber_2011}. The autocorrelation function (ACF) analysis in \texttt{pySYD} then identified a clear peak within the search range, resulting in value of $\Delta\nu=2.307\,\micro$Hz, as we show in Fig.~\ref{fig:acf}. \texttt{PySYD} can estimate uncertainties of both parameters by Monte Carlo-sampling the calculated posterior distributions for the parameters. We discuss the results in Section~\ref{sec:mass} along with the mass estimates.

\subsection{Aliasing in the 2017 SONG Data} \label{sec:song_alias}
As we described above, the RV data starts out in a peculiar manner, dominated by a beating pattern. The peaks in the power spectrum are very close to 1.0, 2.0 and 3.0\,d$^{-1}$, with the 2.0\,d$^{-1}$ (23.15\,$\micro$Hz) peak being the strongest. This could suggest that the true signal is at 2.0\,d$^{-1}$, but daily aliasing makes this conclusion uncertain. We first analysed this beating section of the data set, up until BJD2457685. A classical prewhitening analysis indicates close-by peaks at $2.0037$ and $1.9454$\,d$^{-1}$, creating the beating pattern. Once we removed these, the power spectrum of the residual RV curve displayed a low-amplitude, wide power excess between 1.0 and 4.0\,d$^{-1}$, much more reminiscent of a solar-like oscillator. 

Still, we were left with an ambiguous frequency solution, because a beating of two frequencies at 1.0\,d$^{-1}$ (11.57\,$\micro$Hz) fits the data nearly equally well. However, here we can break this ambiguity thanks to the TESS observations, which place $\nu_{\rm max}$ to $\approx22\,\micro$Hz, as we show in the next Section. Thus, we can assume that the 2.0\,d$^{-1}$ peak is the true signal, and we can proceed with the analysis from there. 

\section{Results} \label{sec:results}
The detection of the oscillation power excess in $\gamma$~Per allows us to estimate the asteroseismic mass of the primary, and to compare it to classical mass inferences. Moreover, with the SONG data at hand we can extend RV seismology into the RGB regime, and investigate how this star relates to already characterized smaller, main sequence (MS) and subgiant (SG) stars.

\subsection{Seismic Parameters and Mass Estimates}
\label{sec:mass}
We calculated mass estimates as described in Section \ref{sec:scale} based on the global seismic parameters we determined. Following the steps described in Section \ref{sec:pysyd}, we derived a pair of $\nu_{\rm max}$ and $\Delta \nu$ values. Analysing the custom light curve, we got $\nu_{\rm max}=21.27 \pm 0.98 \ \micro$Hz and $\Delta \nu = 2.31 \pm 0.10 \ \micro$Hz.
This light curve, as well as its PSD are shown in the two panels of Fig.~\ref{fig:s85-86}. The respective coloured dashed line in the lower panel refers to the derived $\nu_{\rm max}$ value.

\begin{figure}
    \includegraphics[width=0.49\linewidth]{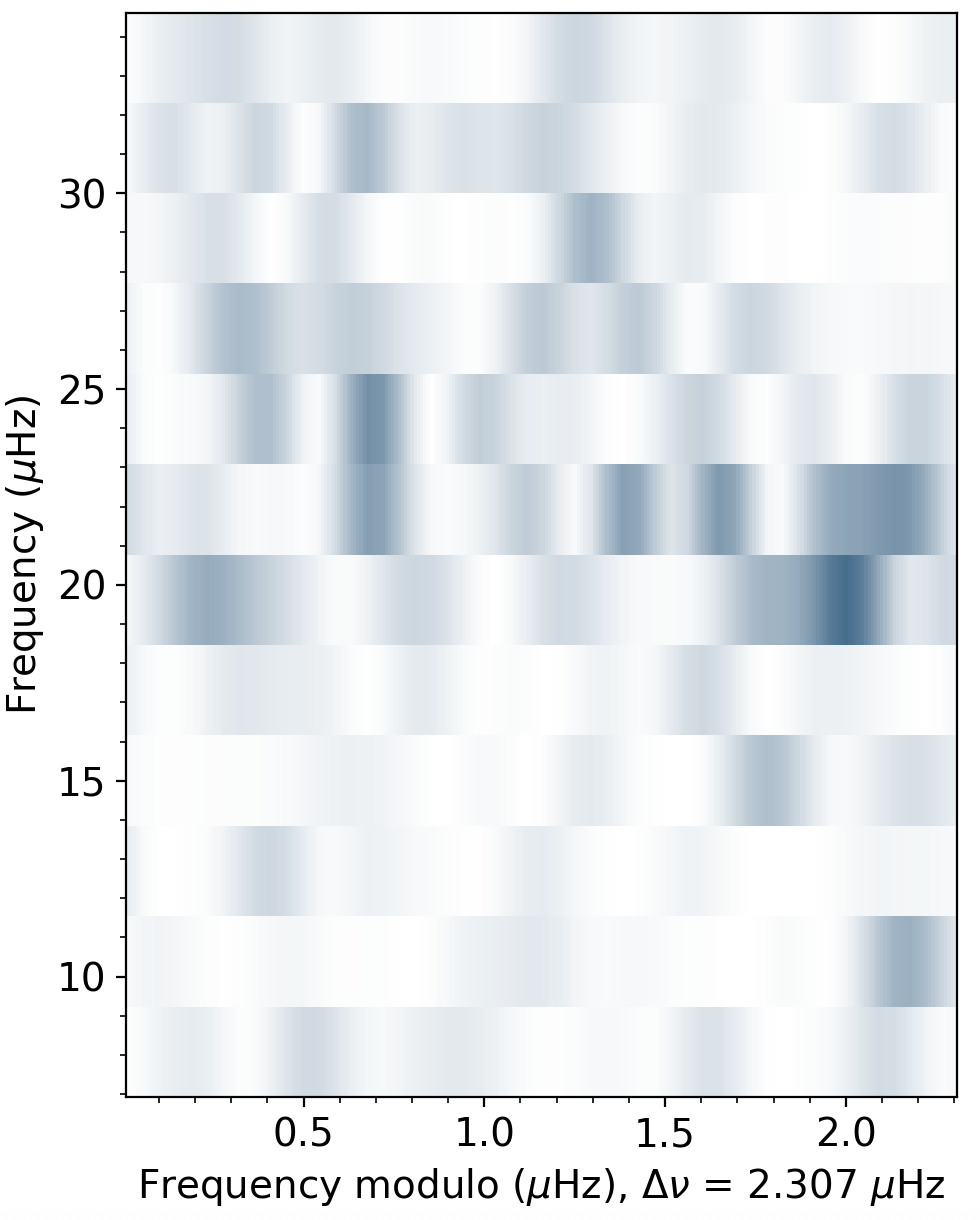}
    \includegraphics[width=0.49\linewidth]{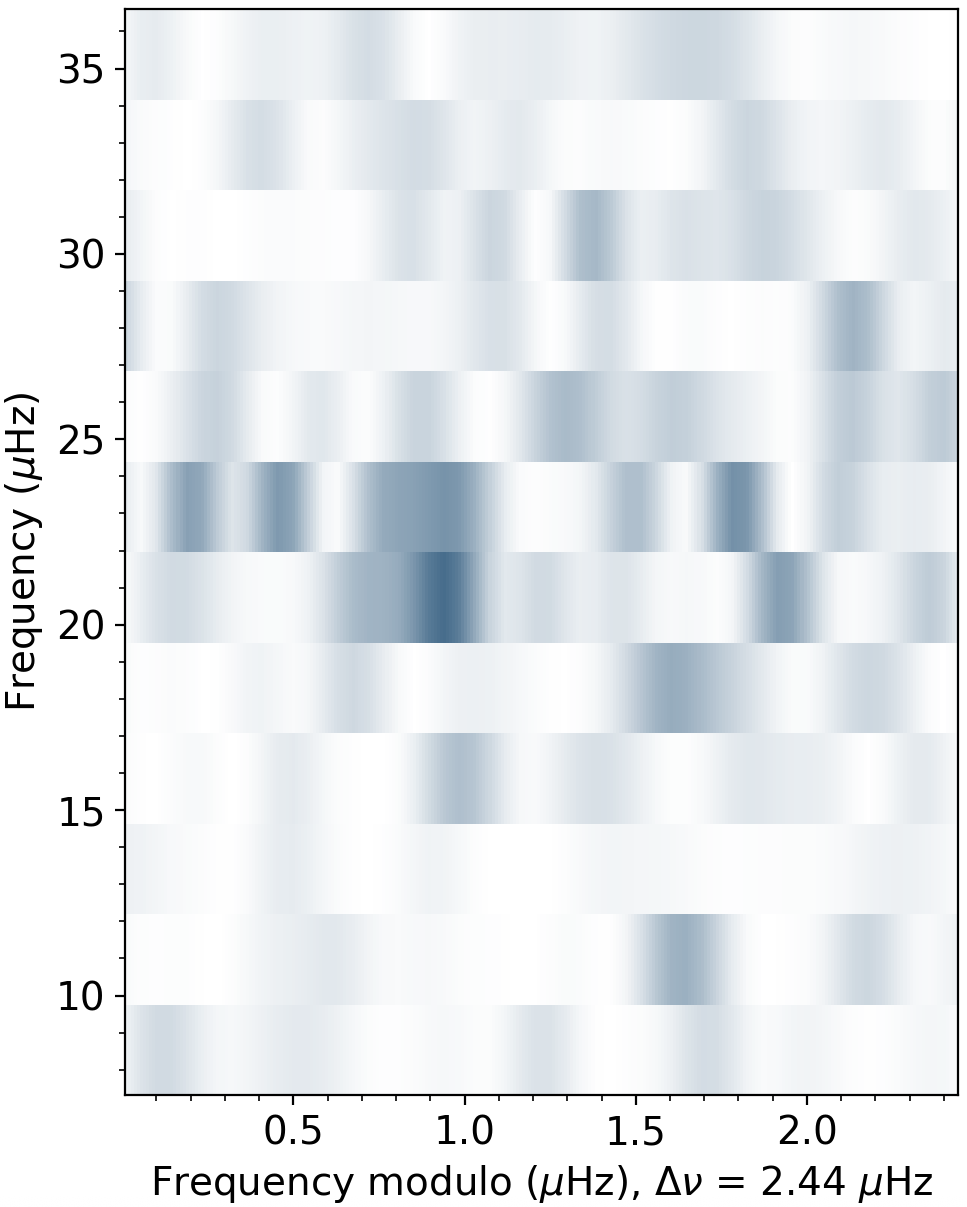}
    \caption{Echelle diagrams of the SNR spectrum of the TESS S85--S86 custom photometry (shown in magenta in Fig.~\ref{fig:s85-86}), using $\Delta \nu = 2.307\,\micro$Hz, based on the \texttt{pySYD} analysis, and $\Delta \nu = 2.44\,\micro$Hz, based on visual inspection of the echelle diagrams, respectively.}
    \label{fig:echelle}
\end{figure}

Based on the $\Delta \nu$ values we obtained, we investigated the echelle diagram of the data using the \texttt{echelle} python package \citep{daniel_hey_2020_3629933}. We found that a value of $\Delta \nu = 2.44\,\micro$Hz gave the best visual result, as shown in the right panel of Fig.~\ref{fig:echelle}. Although the frequency resolution of the data is low, we were able to construct an echelle diagram that shows a repetition of peaks on the left side, at the even mode orders ($\ell=0$ and 2), and a scatter of peaks on the right side, where $\ell=1$ mixed modes are expected. The \texttt{pySYD} $\Delta \nu$ value of 2.307 $\micro$Hz gives a similar result (left panel of Fig.~\ref{fig:echelle}), but with the largest-amplitude peak appearing among the odd modes. KIC~9786910 \citep{Crawford_2024,Crawford_2025} is a similar star to $\gamma$~Per~A with respect to its $T_{\rm eff}$, $\nu_{\rm max}$ and mass, its derived $\Delta \nu$ ($2.384\pm0.035 \ \micro$Hz) is within the common uncertainty ranges of ours, which strengthens our results.

The scatter among the odd modes can be explained by the presence of mixed modes \citep{Beck-2011,Basu-2020,Lindsay-2022}. The $\ell=1$ \textit{p} modes are coupled to $\ell=1$ \textit{g} modes in the core, and this causes deviations from the expected ridge, as weakly coupled modes stay close to the  $\Delta \nu$ frequency spacing, but strongly coupled modes follow the $\Delta \Pi$ period spacing of \textit{g} modes more closely instead. Given the low frequency resolution and the low SNR levels of the PSD, we did not attempt to fit individual peaks in the data, and we did not identify ridges or determined an $\epsilon$ ridge offset parameter, either.

\begin{figure}
    \includegraphics[width=\linewidth]{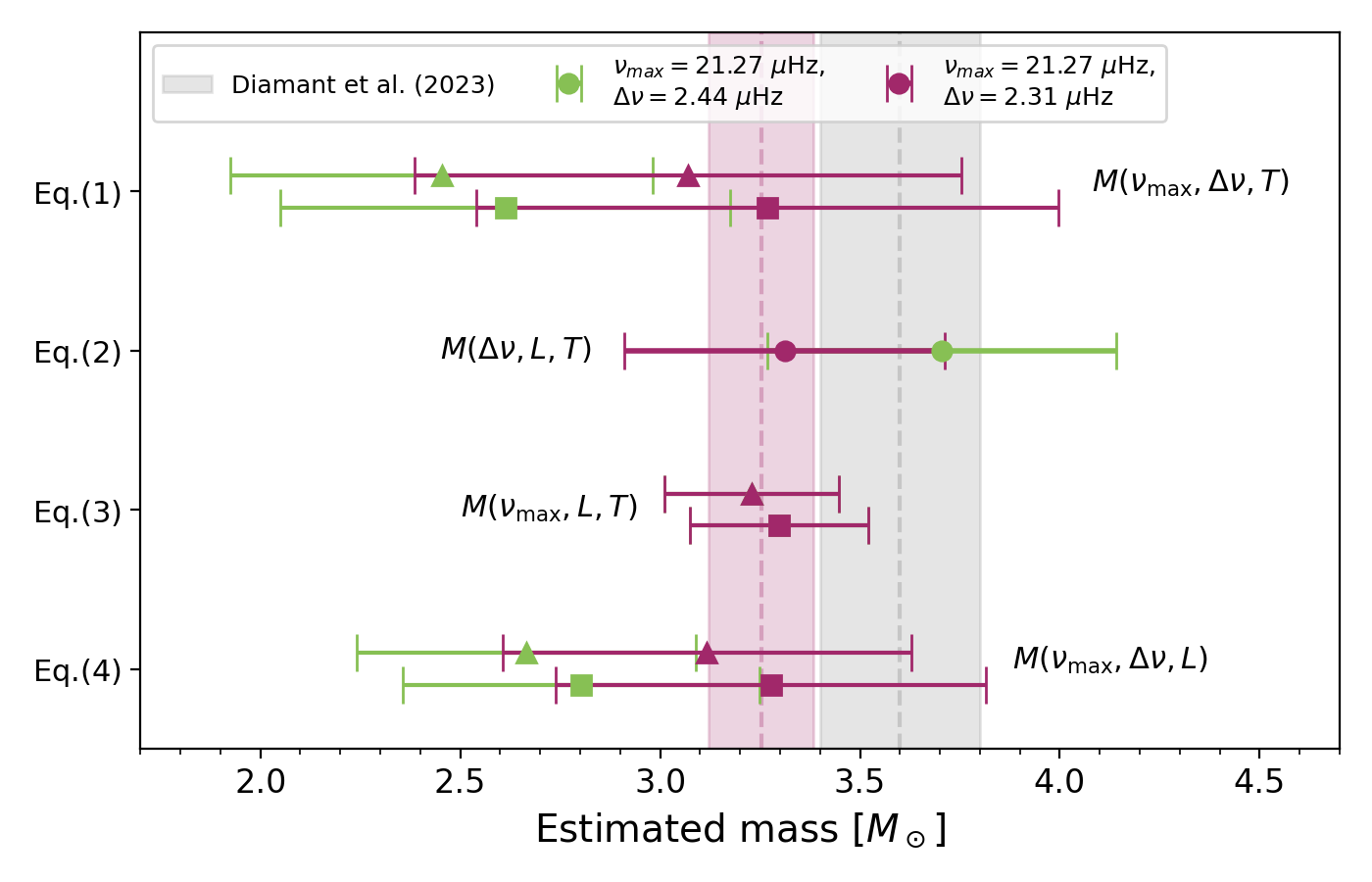}
    \caption{We plotted the results of the mass estimation calculations over the dynamical mass range (grey) published by \cite{Diamant_2023}. Magenta, like earlier, refers to seismic masses determined based on the $\nu_{\rm max}, \Delta\nu$ pair derived from the custom light curve, while for green we used the same $\nu_{\rm max}$ with the visually determined $\Delta\nu=2.44 \ \micro$Hz. Note that Eq.~(\ref{eq3}) uses only $\nu_{\rm max}$ and thus results are identical for both cases. The triangles show the mass calculated with $\nu_{\rm max,\odot}=3141 \ \micro$Hz, while the squares refer to $\nu_{\rm max,\odot} = 3076 \ \micro$Hz. The magenta area displays our estimated mass range.}
    \label{fig:mass_est}
\end{figure}

We used two values for $\nu_{\rm max,\odot}$; $3141 \ \micro$Hz \citep{Andersen_2019} and $3076 \ \micro$Hz \citep{Pinsonneault_2025}, since no consensus has been reached, and it also changes over the solar cycle. We applied $\Delta \nu_\odot = 135.1 \ \micro$Hz based on \cite{Huber_2011}. For the solar temperature we adopted the IAU sanctioned value: $T_\odot = 5772$ K \citep{Prsa_et_2016}. Finally, for $\gamma$ Per A we adopted $T = 4970$ K and $L = 10^{2.45}$ L$_\odot$ from \cite{Diamant_2023}.
The results are in shown in Table~\ref{tab:mas_est} and Fig.~\ref{fig:mass_est}. As in Fig.~\ref{fig:s85-86}, the results based on the custom light curve are shown with orange, while the ones based on the PDCSAP light curve are magenta. The triangles show the calculations with $\nu_{\rm max, \odot}= 3141 \ \micro$Hz, the squares with $\nu_{\rm max, \odot}= 3076 \ \micro$Hz. The grey area refer to the dynamical mass (and its error) derived by \cite{Diamant_2023}.  

The inferred mass values for each scaling relationship are listed in Table~\ref{tab:mas_est}, and shown in Fig.~\ref{fig:mass_est}. We computed the inverse-variance weighted means of the values for both $\Delta \nu$ values. For the \texttt{pySYD} value ($\Delta\nu=2.307 \ \micro$Hz), our result is $M_{\rm seism} = 3.25\pm0.13$\,M$_\odot$, which we accept as the primary result. For the alternate, visually estimated $\Delta \nu$, the same averaging technique gives $M_{\rm seism,  alt} = 3.13 \pm 0.13$ M$_\odot$.

\begin{table*}[ht]
    \renewcommand{\arraystretch}{1.3}
    \centering
    \begin{tabular}{l|c|c|c|c|c|c}
        \multirow{2}{12em}{Seismic parameters} & $\nu_{\rm max,\odot}$ & $M(\nu_{\rm max},\Delta\nu,T)$ & $M(\Delta\nu,L,T)$ & $M(\nu_{\rm max},L,T)$ & $M(\nu_{\rm max},\Delta\nu,L)$ & $\bar{M}$  \\
        & [$\micro$Hz] & [M$_\odot$] & [M$_\odot$] & [M$_\odot$] & [M$_\odot$] & [M$_\odot$] \\ \hline

        \multirow{2}{11em}{$\nu_{\rm max}=21.268\pm 0.98 \ \micro$Hz, $\Delta\nu=2.307\pm0.100 \ \micro$Hz} & $3141$ & $3.07\pm0.68$ & \multirow{2}{5em}{$3.31\pm0.40$} & $3.23\pm0.22$ & $3.11\pm0.51$ & \multirow{2}{5em}{$3.25\pm0.13$} \\
        & $3076$ & $3.27\pm0.73$ & & $3.30\pm0.22$ & $3.28\pm0.54$ & \\ \hline

        \multirow{2}{11em}{$\nu_{\rm max}=21.268\pm 0.98 \ \micro$Hz, $\Delta\nu=2.44\pm0.10 \ \micro$Hz} & $3141$ & $2.45\pm0.53$ & \multirow{2}{5em}{$3.71\pm0.44$} & $3.23\pm0.22$ & $2.66\pm0.42$ & \multirow{2}{5em}{$3.13\pm0.13$} \\
        & $3076$ & $2.61\pm0.56$ & & $3.30\pm0.22$ & $2.80\pm0.45$ & \\
    \end{tabular} \vspace{0.1cm}
    \caption{Seismic mass estimations (with errors) derived from four different scaling relations using the $\nu_{\rm max}$ determined from our custom light curve. In the first line $\Delta\nu$ is also derived from our custom light curve, while its value in the second line is based on visual inspection of the echelle diagrams (see Fig.~\ref{fig:echelle}).}
    \label{tab:mas_est}
\end{table*}

\subsection{Photometric Amplitudes} \label{sec:phot_amp}
Solar-like oscillations are stochastic by nature, so individual mode amplitudes vary over time -- as the 2017 SONG data clearly demonstrates. Therefore, if we want to characterize the strength of the oscillations, we need to calculate averaged amplitudes per oscillation modes, as defined by \citet{Kjeldsen_2008}. 

The first steps in their method is to heavily smooth the power excess in the PSD with a Gaussian filter and to remove the background signal coming from granulation noise. We followed \citet{Kjeldsen_2008} and set  the FWHM (full width at half maximum) of the Gaussian to be $4\Delta\nu$ for the smoothing. Since the power excess has a low SNR, we used multiple methods to test the robustness of our amplitude values.  
We fitted the granulation background in the TESS data with multiple Harvey-functions using MCMC, while excluding the modes themselves within the region of the power excess. 
After removing the background signal, we smoothed the PSD with the sliding Gaussian filter described above. The two different background fits resulted in small differences and gave us additional handles on the amplitude uncertainties. This type of processing is similar to the way $\nu_{\rm max}$ is calculated in \texttt{pySYD}, and resulted in a very similar value of $21.1\,\micro$Hz.

The average amplitude is calculated from the peak value of the smoothed PSD. In the last step, \citet{Kjeldsen_2008} multiplies this value with $\Delta\nu/c$ to obtain the average amplitude per oscillation mode, where $c$ is a factor that measures the effective number of modes per order. The value of $c$ depends on the sensitivity of the observations to the various low-degree modes. This sensitivity will be different both for the method we use (photometry or RVs), and for the wavelength range of the photometric observations, as well. For TESS observations, we used $c=2.95$ \citep{Kjeldsen_2008}. 

In the case of $\gamma$~Per, we also have to account for the binary nature of the star. Based on the luminosities published by \citet{Diamant_2023}, we calculate the dilution factor of the luminosity of $\gamma$~Per~A to be $(L_{\rm A} + L_{\rm B})/L_{\rm A} = 1.24$. Finally, since we are interested in the bolometric amplitudes, we also calculated the bolometric correction factor for the luminosity to be $c_{\rm P-bol} = 1.0395$, as defined by \citet{Lund_2019} for the equation $A_{\rm bol} = c_{\rm P-bol}\,A_{\rm P}$. With these corrections implemented, we arrive to an average bolometric luminosity amplitude per oscillation mode of $A_{\rm bol} = 36.6 \pm 1.5$\,ppm.

Originally, \citet{Kjeldsen-Bedding_1995} derived that oscillation amplitudes scale by $A_{\rm bol} = (L/M)\,(T_{\rm eff,\odot}/T_{\rm eff})\, A_{\rm bol,\odot}$. This means that higher-mass red giants at the same luminosity (or $\nu_{\rm max}$) level will have lower amplitudes, which is nicely illustrated if we compare $\gamma$~Per to a similar but lower-mass star. For HD\,145250, for example, even single-sector observations provided stronger power excess detections, despite being a fainter target \citep{Molnar-2025RNAAS}. However, the scaling relation of \citet{Kjeldsen-Bedding_1995} would predict amplitudes on the order of $\approx300$\,ppm for $\gamma$~Per~A, which is still an order of magnitude higher than what we see with TESS. 

Since we cannot model mode amplitudes for oscillations numerically, validating the scaling relation requires comparison of observed amplitudes with physical parameters of real stars. This has been investigated multiple times, mostly based on \textit{Kepler} observations. \citet{Stello_2011}, for example, compared the mode amplitudes of three open clusters within the original \textit{Kepler} field-of-view (FoV) to estimates based on isochrones fitted to the clusters, adding exponents to the $L/M$ and $T_{\rm eff}$ scaling. They found a good agreement between the observed and predicted slopes of the $A_{\rm bol}$--$\nu_{\rm max}$ relation, but the observations revealed a clear mass-related separation, which the scaled models did not reproduce. Based on a larger sample of field stars in the \textit{Kepler} FoV, \citet{Huber_2011} further refined the amplitude scaling, adding separate exponents to $L$ and $M$, to account for the observed stronger mass dependence.

\begin{figure}
    \centering
    \includegraphics[width=\linewidth]{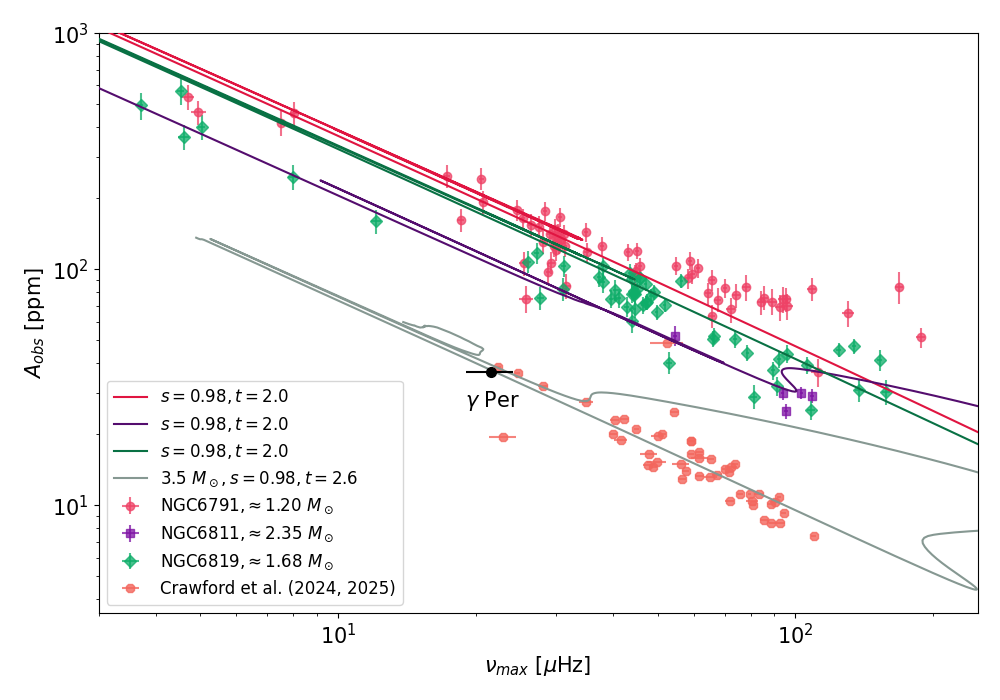}
    \caption{
    Comparison of the measured average mode amplitudes between $\gamma$~Per~A (black dot) and the three open clusters in the original \textit{Kepler} FoV (pink, purple and green dots) by \citet{Stello_2011}, and other high-mass \textit{Kepler} stars (orange) published by \cite{Crawford_2024} with updated masses from \cite{Crawford_2025}. Models with lines in respective colours of the open clusters and grey from MIST.} 
    \label{fig:phot_amp_compare}
\end{figure}

We plot the open cluster data from \citet{Stello_2011} in Fig.~\ref{fig:phot_amp_compare}, with the estimated average masses indicated, and along with with $A_{\rm bol}$--$\nu_{\rm max}$ relations based on MIST isochrones (MESA Isochrones and Stellar Tracks, \citealt{Choi-2016,Dotter-2016}) interpolated to the ages and metallicities of the clusters. We tested the scaling defined by \citet{Huber_2011} and found that the $L$ and $M$ scaling exponents $s=0.98$ and $t=2.0$ reproduce the cluster observations the closest.
We then added oscillation amplitudes published by \cite{Crawford_2024,Crawford_2025}, who specifically targeted heavier red giants above $2.5$ M$_\odot$ in the \textit{Kepler} field.
We also included a stellar track at $M=3.5$ M$_\odot$ with modified scaling exponents of $s=0.98$ and $t=2.6$ that align with $\gamma$~Per~A and the higher-mass \textit{Kepler} stars. Fig.~\ref{fig:phot_amp_compare} clearly shows that the observed amplitude of the star, as well as all the other high-mass stars, are significantly lower than the predicted value based on the mass scaling relation of \citet{Kjeldsen-Bedding_1995}, in agreement with the conclusions of \citet{Crawford_2024}.  
This further suggests that the scaling relations need to be modified to describe the amplitudes of more massive stars accurately.

\subsection{Oscillation Amplitudes from RVs} \label{sec:rv_amp}
We also calculated the average amplitude per oscillation mode from the RV data. We again followed the method of \citet{Kjeldsen_2008}. Unlike our earlier photometric analysis, here we did not need to account for the binarity. We used an effective mode number coefficient of $c = 4.09$. For the RVs we used two different approaches to fit the granulation: we fitted both a single Harvey-function, and the multi-Harvey MCMC algorithm. This was necessary as the multi-component fit struggled with converging to a good solution in some cases. This way we could obtain an upper and a lower limit for the amplitudes. This step was followed by the same smoothing process as described in Section~\ref{sec:phot_amp}. The main deviation from the photometry is that we used the PSDs of the GP models instead of the PSDs of the actual RV data. We did this to prevent aliasing from influencing our calculations. As we described in Section~\ref{sec:song_data}, the GP model PSDs reproduce the oscillation signals well. 

\begin{figure}
    \includegraphics[width=\linewidth]{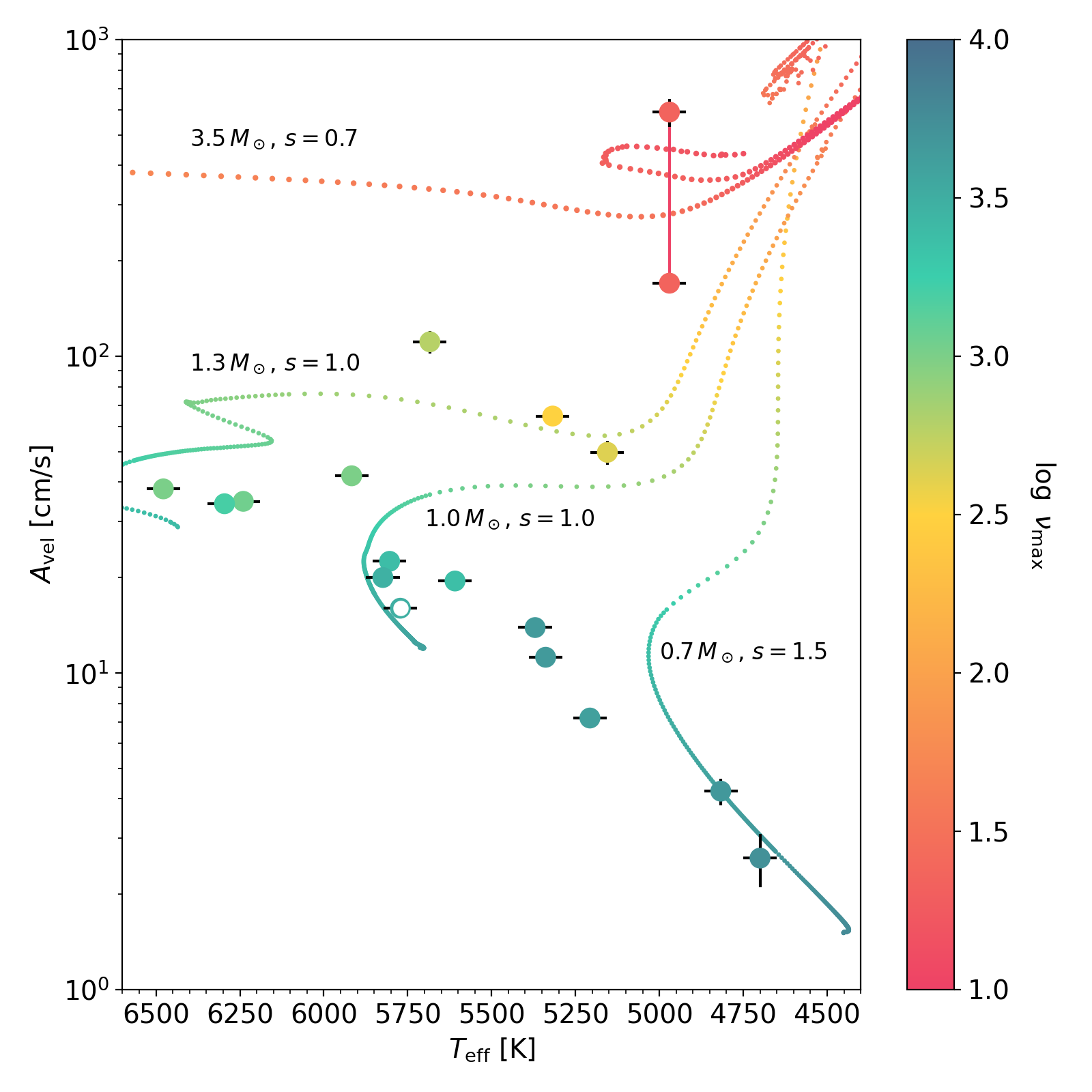}
    \caption{The distribution of $A_{\rm vel}$ amplitudes against $T_{\rm eff}$, colour-coded by $\log \nu_{\rm max}$. The evolutionary tracks follow four different initial stellar masses, and change colour according to their $\log \nu_{\rm max}$. The exponents we used in the $A_{\rm vel}$ scaling relations are also indicated.
    The HRD is clearly visible with most of the stars forming the MS, three stars falling onto the SG branch ($\beta$~Aql, $\nu$~Ind, HD35833), and $\gamma$~Per~A falling on the RGB or RC. The Sun is marked with an empty circle.}
    \label{fig:comp_teff}
\end{figure}

\begin{figure}
    \includegraphics[width=\linewidth]{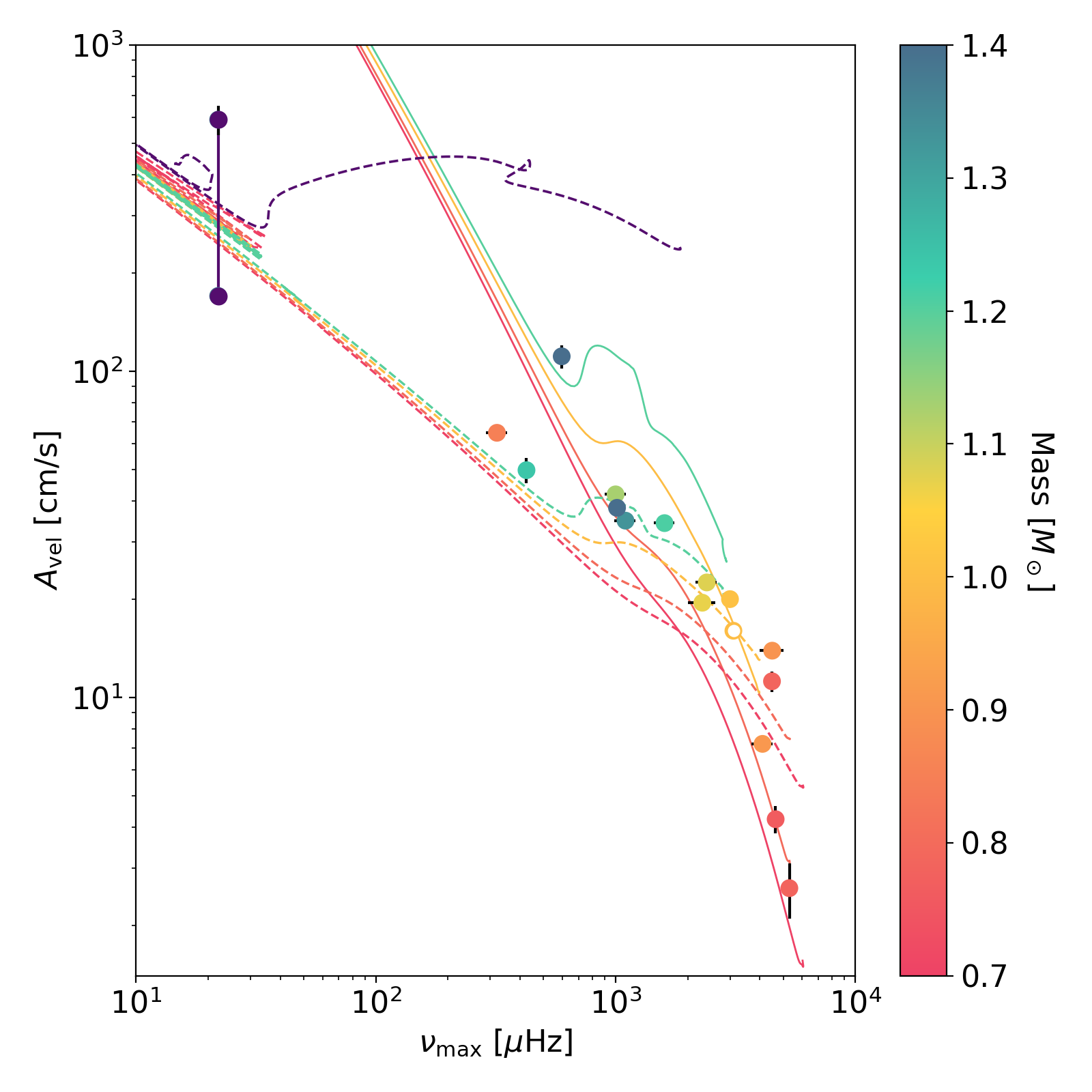}
    \caption{Same comparison as in Fig.~\ref{fig:comp_teff} on the $\nu_{\rm max}$--$A_{\rm vel}$ plane, this time colour-coded with stellar mass. Here, the same coloured evolutionary tracks differ with amplitude scaling, dashed lines refer to $(L/M)^{0.7}$, while solid lines $(L/M)^{1.5}$. Sun is shown with an empty circle.}
    \label{fig:comp_numax}
\end{figure} 

The fit for the GP model of the 2024 data is shown in the upper section of Fig.~\ref{fig:song_2024}; the 2017 fits are shown in the Appendix in two panels of Fig.~\ref{fig:song_2017_dense}. Since the 2017 data is largely dominated by only a few modes, we did not consider the $\nu_{\rm max}$ values for those segments. For the 2024 data, we obtained $\nu_{\rm max}=22.5\pm1.0\,\micro$Hz, which is slightly higher than the photometric value, and the resulting seismic mass would align better with the dynamical mass. Still, given the sparseness of the observational data behind the GP models, we consider the photometric $\nu_{\rm max}$ to be our primary result.

Based on these fits, the average RV amplitude of $\gamma$~Per~A is between $170\pm30$ (2024 data) and $590\pm60$~m/s (first segment of the 2017 data). The various RV amplitudes are listed in Table~\ref{tab:song_rv}, which were derived from their PSDs as shown in Figs.~\ref{fig:song_2024}, \ref{fig:song_2017_densea_backfits}, \ref{fig:song_2017_denseb_backfits}, \ref{fig:song_2017_sparse_p1b_backfits}, \ref{fig:song_2017_sparse_p1a_backfits} and \ref{fig:song_2017_sparse_p2_backfits}. The average RV amplitude is significantly higher than the RV amplitudes measured for MS and SG stars which are in the 0.02--0.65 m/s range \citep{Kjeldsen_2008,Campante-2024}. Our observation then leads to an important question: which amplitude scaling relation extends into the RGB/red clump (RC) regime correctly? \citet{Kjeldsen-Bedding_1995} originally proposed a simple $A_{\rm vel} \propto (L/M)^s$ relation with $s=1.0$. Theoretical models and observations put the exponent to $s=0.7$--$1.5$ \citep{Houdek-1999,Samadi-2005}, with observations favouring the higher end of the range \citep{Campante-2024}. 

We compiled a list of stars for which oscillation parameters ($\nu_{\rm max}$, $\Delta \nu$, $A_{\rm vel}$), $T_{\rm eff}$, and mass values have been published (see Table~\ref{tab:stars_data}). Where only approximate values were provided for $\nu_{\rm max}$, we assumed a 10\% uncertainty. We show the distribution of $A_{\rm vel}$ amplitudes against $T_{\rm eff}$, colour-coded by $\log \nu_{\rm max}$ in Fig.~\ref{fig:comp_teff}. The distribution of the points clearly reproduce the HRD, with most of the stars forming the MS, three stars falling onto the SG branch ($\beta$~Aql, $\nu$~Ind, HD35833), and $\gamma$~Per~A falling on the RGB. $\gamma$~Per~A is shown with two dots connected by a line, as we determined two values for its $A_{\rm vel}$ from 2017 and 2024, respectively (Sect.~\ref{sec:rv_amp}). We plot also includes four stellar evolutionary tracks from the MIST database. Three of these show theoretical $T_{\rm eff}$ and $A_{\rm vel}$ values for non-rotating, solar-metallicity models with masses of $M = 0.7$, $1.0$, and $1.3$ M$_\odot$. The fourth line is a stellar track representing $\gamma$~Per~A, at $M = 3.5$ M$_\odot$ and [Fe/H] = $-0.2$. As noted in the plot, we used different $s$ values as well, between $0.7$ and $1.5$.

We then plotted the same stars onto the $\nu_{\rm max}$--$A_{\rm vel}$ plane in Fig.~\ref{fig:comp_numax}, this time colour-coded with stellar mass. This plot also includes a set of stellar evolutionary tracks from the MIST database. Four tracks show theoretical $\nu_{\rm max}$ and $A_{\rm vel}$ values for non-rotating, solar-metallicity models with masses of $M = 0.7$, $0.8$, $1.0$, and $1.2$ M$_\odot$. The fifth line is a stellar track representing $\gamma$~Per~A, at $M = 3.5$ M$_\odot$ and [Fe/H] = $-0.2$. We calculate $\nu_{\rm max}$ from Eq.~(\ref{eq3}), and the RV amplitudes as:
\begin{equation}
    A_{\rm vel} = (L/M)^s\, A_{\rm vel,\odot}.
\end{equation}
Two sets of lines show the difference between the two extrema of the amplitude scaling, at $s=0.7$ (dashed lines) and 1.5 (solid lines) in Fig.~\ref{fig:comp_numax}. 

As the plot shows, stars near the Sun (displayed with a diamond) can be fitted with either exponent reasonably well. Away from the Sun, differences become clearer. The K dwarf stars $\epsilon$~Ind and HD219134, and the SG star HD35833 can only be fitted by the highest, $s=1.5$ exponent. Our result on $\epsilon$~Ind is in agreement with the findings of \citet{Campante-2024}. In contrast, higher-mass MS stars, like Procyon, the other two SG stars, $\beta$~Aql~A and $\nu$~Ind, as well as $\gamma$~Per~A require lower exponents. For the $3.5$ M$_\odot$ track in particular (purple dashed line), we only show the $0.7$ scaling, because if we use the $1.5$ exponent the values shift into the $\sim 10^5$\,cm/s range, and run well outside of the plotted area. Unfortunately, \citet{Knudstrup-2023} did not publish the amplitudes for $\gamma$~Cep~A, which would probe the split between exponents further at $\nu_{\rm max}=186\,\micro$Hz. Our results indicate that the exponent of the RV amplitude scaling should feature another parameter dependence, such as $\log g$ or $R$, to account for the differences in slope for non-solar-like stars.

\begin{table}[h!]
    \renewcommand{\arraystretch}{1.3}
    \centering
    \caption{The RV amplitude per oscillation mode values we determined based on GP fits with different quality factors applied to specific data sections.}    \label{tab:song_rv}
    \begin{tabular}{l c}
        \hline 
        Data segment & $A_{\rm vel}$ [cm/s] \\ 
        \hline
        2017 dense data, $Q = 9$ & $590\pm60$ \\
        2017 dense data, $Q = 10$ & $470\pm 50$ \\
        2017 sparse data, part 1, $Q = 18$ & $420\pm60$ \\
        2017 sparse data, part 1, $Q = 20$ & $310\pm 50$ \\
        2017 sparse data, part 2, $Q=20$ & $210\pm30$ \\ \hline
        2024 data, $Q=10$ & $170 \pm30$ \\
        \hline
    \end{tabular}
\end{table}

\subsection{Ratio of Photometric and RV Amplitudes} \label{sec:ratio}
Just like the frequency parameters of the oscillations, $\nu_{\rm max}$ and $\Delta \nu$, scale with the physical parameters of stars, so do the oscillation amplitudes. Scaling relations for intensity and RV amplitudes were proposed by \citet{Kjeldsen-Bedding_1995}, and were refined in their later work \citep{Kjeldsen-Bedding_2011}. These relations predict that the ratio of the bolometric intensity and RV amplitudes scale inversely with the effective temperature of the star:  $A_{\rm bol}/\varv_{\rm osc}~\sim~T_{\odot}/T_{\rm eff}$, which can be tested against the solar value.

\begin{figure}
    \includegraphics[width=\linewidth]{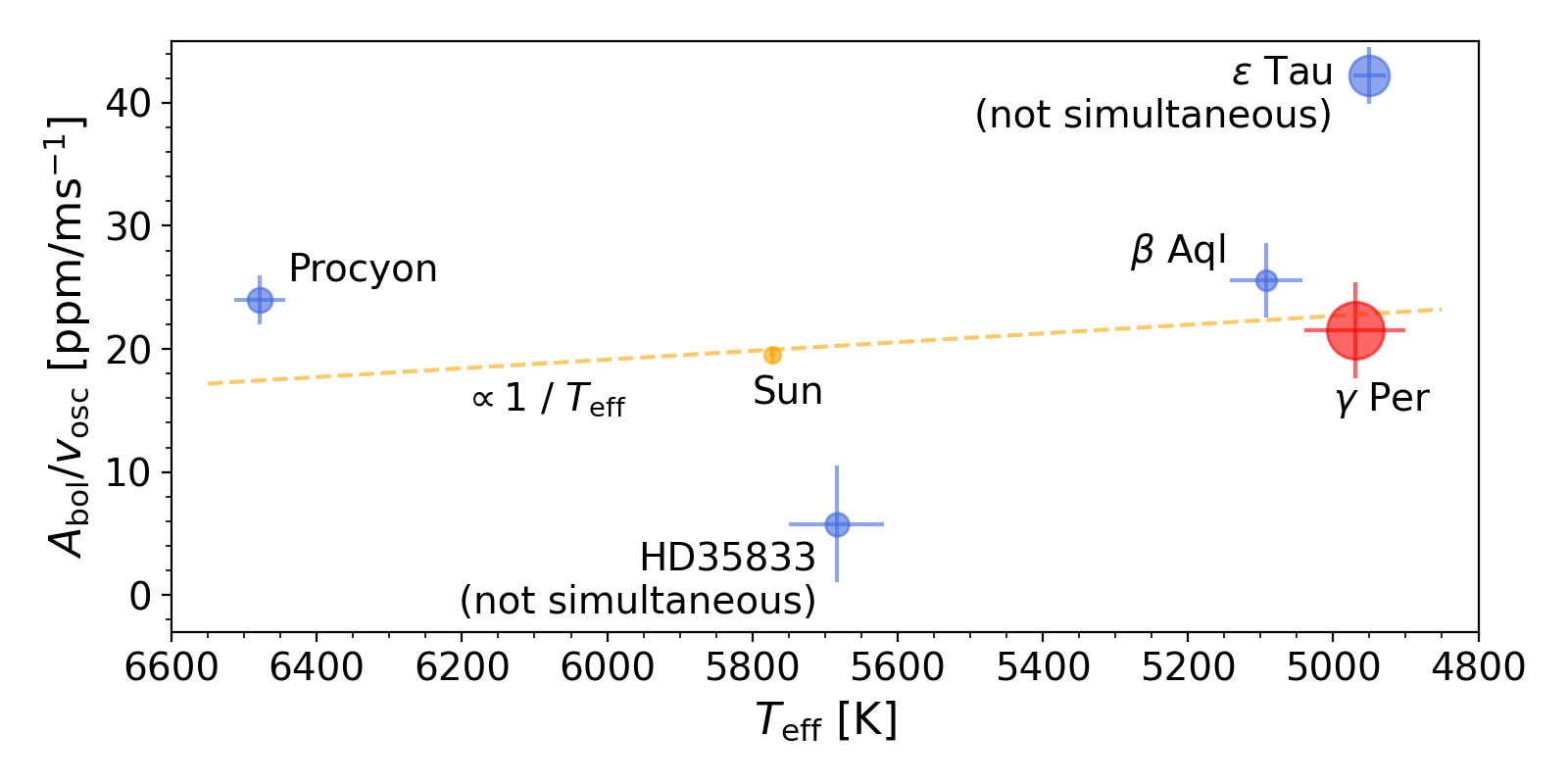}
    \caption{We determined the average oscillation amplitude per mode quantities $A_{\rm bol}$ and $\varv_{\rm osc}$ for the light curve and RV data following \cite{Kjeldsen_2008}. We find that their ratio follows the expected $1/T_{\rm eff}$ scaling as predicted by \cite{Kjeldsen-Bedding_1995}. Here, we show a comparison with two other stars. Point sizes correspond to stellar mass.}
    \label{fig:amp_ratio}
\end{figure}

Since the 2024 TESS and SONG observations were taken contemporaneously, we can calculate the RV to luminosity amplitude ratio for $\gamma$~Per. The solar ratio is $A_{\rm bol,\odot}/A_{\rm vel,\odot} = 19.5 \pm 0.7$\,ppm/ms$^{-1}$ \citep{Kjeldsen-2025}. For $\gamma$~Per we calculate this ratio to be $A_{\rm bol}/v_{\rm bol} = 21.5 \pm 0.2$\,ppm/ms$^{-1}$. The ratio between the solar and $\gamma$~Per values is $1.10 \pm 0.04$. Since the (inverse) ratio of the effective temperature of $\gamma$~Per against the Sun is $1.16\pm0.02$, and we expect the oscillation amplitude ratios to scale with that, as Fig.~\ref{fig:amp_ratio} shows, the $A_{\rm bol}/A_{\rm vel}$ value we got agrees with the expected value within $1.4\sigma$.

We also find good agreement with the amplitude ratio of $\beta$~Aql, based on the results of \citet{Kjeldsen-2025}. Another star, for which photometric and RV oscillations have been determined in the same way, is HD35833 \citep{Gupta_2022}. This star is offset from the expected ratio, but these measurements were not contemporaneous and could indicate a change in oscillation amplitudes over time, as well. Strengthening the idea that observations that are not contemporaneous yield different amplitude ratios, the non-simultaneous observations of $\epsilon$~Tau by \cite{Arentoft_2019} also show a similar offset. To state anything more further investigation is needed, and these cases prove that this would be an interesting area of research.
More than a decade ago, \cite{Huber_2011_2} derived a similar ratio ($24 \pm 2$ ppm/ms$^{-1}$) for Procyon A based on \textit{MOST} and ground-based RV observations.
So far, Fig.~\ref{fig:amp_ratio} suggests that stars cooler than the Sun follow the expected ratio relatively well, while Procyon, a hotter star, does not. However, we are in a dire need of further observations to draw any conclusions.

\section{Discussion \& Conclusions} \label{sec:discussion}
In this paper we carried out the first asteroseismic analysis of the bright RGB primary component of the $\gamma$~Persei system, based on photometry collected by TESS, and infer a seismic mass of $3.25\pm0.13$ M$_\odot$.
We also detected the oscillations in RVs from measurements collected with SONG.
Sadly, even after obtaining simultaneous observations by TESS and SONG from 2024, we cannot compare them directly, as both are too sparse, especially the SONG data. Despite, we were able to determine the mode-averaged oscillations amplitudes both from photometry and from RVs.

After comparing our mass and amplitude values with literature data, we can state that $\gamma$ Per A clearly does not fit the results for open clusters published by \cite{Stello_2011} in Fig.~\ref{fig:phot_amp_compare}; the photometric amplitude appears to be too low. However, it perfectly aligns with the high-mass stars published by \cite{Crawford_2024, Crawford_2025}. Therefore, we conclude that the amplitude scaling must have a much steeper mass dependence at higher masses, possibly with $t=2.6$. Fast rotation and strong magnetic activity can inhibit solar-like oscillations \citep{Chaplin_2011,Gaulme_2014,Mathur-2019}, and simulations show that mergers can create highly magnetic massive blue straggler stars \citep{Schneider-2019}. However, they subsequently spin down, and their magnetic fields may weaken by the time they evolve to the RGB. Being a likely MS-MS merger product \citep{TND-2026} seems to have no effect on its own on the photometric oscillation amplitudes of $\gamma$~Per as it fits perfectly among other oscillating high-mass red giants. Resolving the conundrum of low amplitudes in massive RGB stars therefore will require further observations, including magnetic field strength or chromospheric activity indicator measurements \citep{Gehan-2024}.

Considering the HRD formed by stars with RV oscillation amplitude data in Fig.~\ref{fig:comp_teff}, $\gamma$ Per falls right where expected, on the RGB. Thus, we could argue that we are in the right range of $A_{\rm vel}$ value. We also note that according to our findings, a different exponent needed for red giants regarding the $A_{\rm vel}$ scaling relation to MS and SG stars.

Studying the $\nu_{\rm max}$--$A_{\rm vel}$ plane in Fig.~\ref{fig:comp_numax}, we see that $\gamma$ Per prefers the evolutionary track scaled with $(L/M)^{0.7}$, and aligns even with low-mass stars. We do not have enough comparison stars to argue for either set of tracks. Most of them, just like our red giant, follow the dashed lines and $s=0.7$ scaling. On the other hand, the lowest and highest (other than $\gamma$ Per) mass stars suggest evolutionary tracks scaled with $(L/M)^{1.5}$. As the study of evolutionary tracks is not scope of this paper, we shall not go into further detail -- though, we note it is a matter worth investigating.

As discussed in Section~\ref{sec:ratio}, our results of $\gamma$ Per are in good agreement with the expected $1/T_{\rm eff}$ scaling of the photometric and RV amplitude ratio. Though, Fig.~\ref{fig:amp_ratio} is in dire need of further data points, ergo stars for which both $A_{\rm bol}$ and $v_{\rm osc}$ are determined, preferably contemporaneously, in order to argue the validity of this relation.

Overall, $\gamma$~Per highlights the importance of studying high-mass red giant stars, and validates the effort needed to collect RV data for red giant oscillators. While several such stars are potentially accessible by TESS, RV observations require more extensive efforts. One target in particular which would be amenable to a similar analysis is $\alpha$~UMa~A, which is brighter and appears to be of similar mass as $\gamma$~Per A \citep{Buzasi-2000,Rudrasingam-2026}.

\begin{acknowledgements}
R.~Z.~Á. and L.~M. would like to thank Daniel Hey and Benjamin Pope for discussions on Gaussian Processes fitting. R.~Z.~Á. would also like to thank the ``SONG Review Publication Panel'', moreover Mikkel Lund and Dennis Stello for their insightful comments about the manuscript.  Cs.~K. would like to thank Madeline Howell for her help with implementing the linear background fitting model in \texttt{pySYD} and for providing a more suitable approach to estimate global seismic parameters and their their uncertainties for low-frequency, low-SNR stars. Cs.~K. would also like to thank Oliver Hall for his help in creating an automated MCMC background fitting model for analysing low-amplitude solar-like oscillators.
This research was supported by the `SeismoLab' KKP-137523 \'Elvonal grant and the NKFIH excellence grant TKP2021-NKTA-64 of the Hungarian Research, Development and Innovation Office (NKFIH). This project received funding from the LP2025-14/2025 Lendület grant of the Hungarian Academy of Sciences. This paper includes data collected with the TESS mission, obtained from the MAST data archive at the Space Telescope Science Institute (STScI). Funding for the TESS mission is provided by the NASA Explorer Program. STScI is operated by the Association of Universities for Research in Astronomy, Inc., under NASA contract NAS 5–26555. 
This publication includes observations made with the SONG network of telescopes operated by Aarhus University, Instituto de Astrofísica de Canarias, the National Astronomical Observatories of China, University of Southern Queensland and New Mexico State University.
This research made use of NASA’s Astrophysics Data System Bibliographic Services, as well as of the SIMBAD and VizieR databases operated at CDS, Strasbourg, France.
\end{acknowledgements}

\bibliographystyle{aa}
\bibliography{references}


\begin{appendix}

\section{TESS Sector 58 PSD and S85--86 Custom Light Curve}
\begin{figure*}[h!]
    \includegraphics[width=\linewidth]{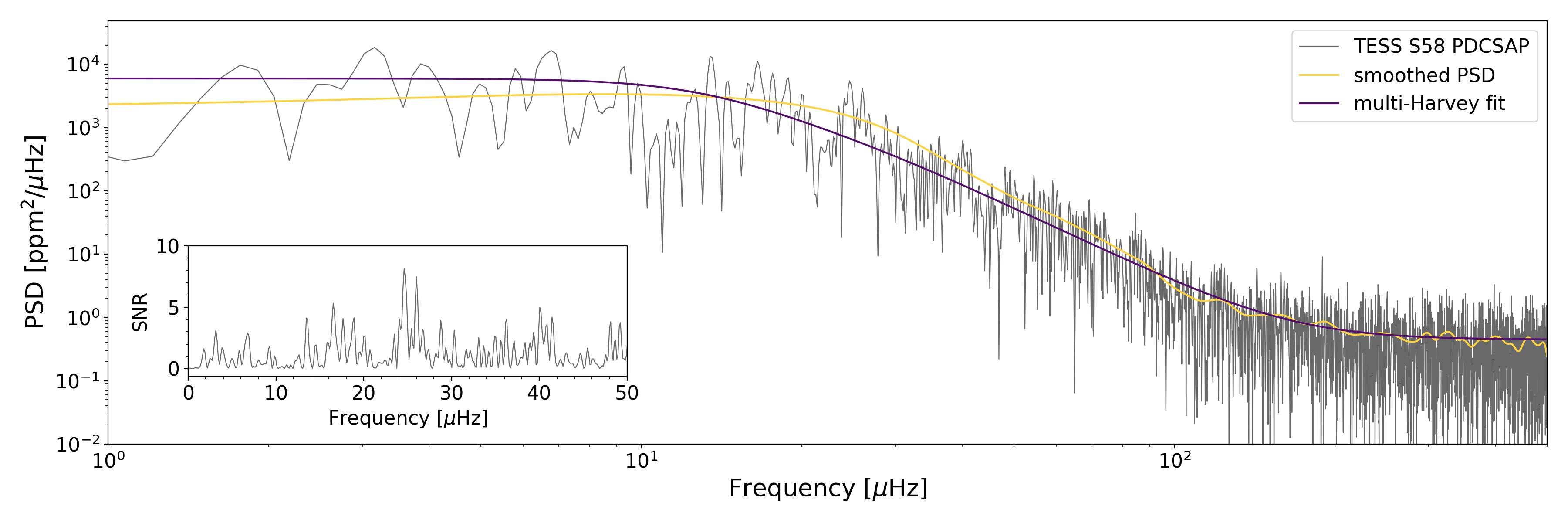}
    \caption{The PSD of TESS Sector 58 PDCSAP light curve is shown with grey, the smoothed version is plotted over with yellow, the multi-component Harvey-like functions fit is with purple. The inset displays the SNR of the PSD.}
    \label{fig:tess_s58_psd}
\end{figure*}

To include as many TESS sectors as possible, in this section we present the PSD of TESS S58 PDCSAP light curve (see Fig.~\ref{fig:tess_s58_psd}), which we discuss in Section~\ref{sec:pysyd}. The inset of plot shows the mentioned power excess.

The first 10 lines of our custom light curve for TESS S85--86 is in Table~\ref{tab:lc_tess8586}, which is shown with orange in the upper panel of Fig.~\ref{fig:s85-86}.

\begin{table}[h!]
    \renewcommand{\arraystretch}{1.3}
    \centering
    \caption{Sample table of the custom light curve of TESS S85--86. The full table is available in machine-readable format in the electronic version.}
    \label{tab:lc_tess8586}
    \begin{tabular}{l l l}
        \hline 
        BJD--2450000 & Normalised flux & Flux error\\
        (days) & (ppm) & (ppm) \\
        \hline
        $3610.559580$ & $1000270.395155$ & $23.905707$ \\
        $3610.560969$ & $1000257.896040$ & $23.905972$ \\
        $3610.562358$ & $1000291.292855$ & $23.905575$ \\
        $3610.563747$ & $1000289.284790$ & $23.905590$ \\
        $3610.565136$ & $1000300.270852$ & $23.906213$ \\
        $3610.566525$ & $1000273.825517$ & $23.906836$ \\
        $3610.567914$ & $1000363.728699$ & $23.907461$ \\
        $3610.569303$ & $1000315.707129$ & $23.907508$ \\
        $3610.570692$ & $1000321.091593$ & $23.906256$ \\
        $3610.572080$ & $1000321.231168$ & $23.906539$ \\
        \multicolumn{3}{l}{...}\\
        \hline
    \end{tabular}
\end{table}

\section{TESS halo photometry}
Saturated stars have represented a unique problem for the Kepler/K2 and TESS missions. While the detectors have been able to preserve flux via bleeding excess electrons along the CCD columns, summing that flux up was not always straightforward. Saturation columns could reach the edges of the CCD modules, could be contaminated by other stars, and, in the case of imagette data, could have exceeded the predefined imagette edges. Originally developed for the K2 mission, halo photometry works with the unsaturated pixels of the imaging halo around the target. These pixels are identified, summed, and then weights for each pixel are optimized by minimizing a total variation function, and thus minimizing the non-astrophysical jumps between data points \citep{White-2017,Pope-2019}.

\begin{figure*}[h!]
    \centering
    \includegraphics[width=\linewidth]{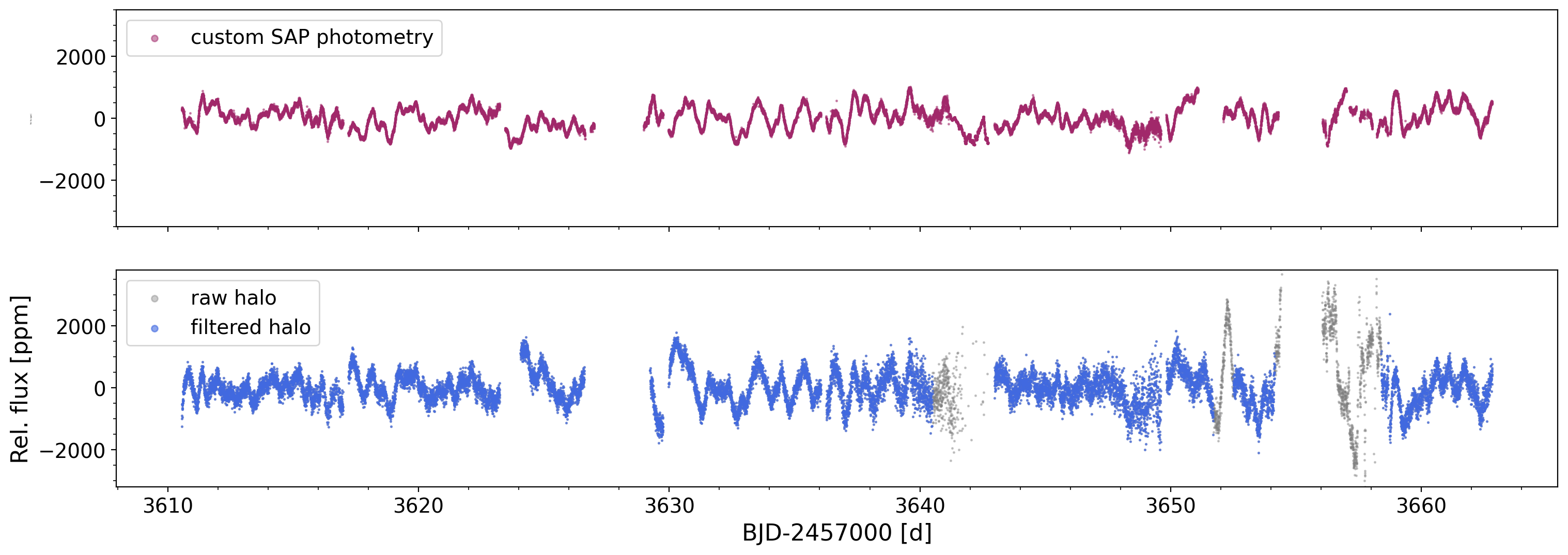}
    \caption{\textit{Upper panel}: our Sectors 85--86 custom SAP light curve of $\gamma$~Per, as shown in Fig.~\ref{fig:s85-86}. \textit{Lower panel}: halo photometry of Sectors 85--86.}
    \label{fig:halo_lc}
\end{figure*}

\begin{figure*}[h!]
    \centering
    \includegraphics[width=\linewidth]{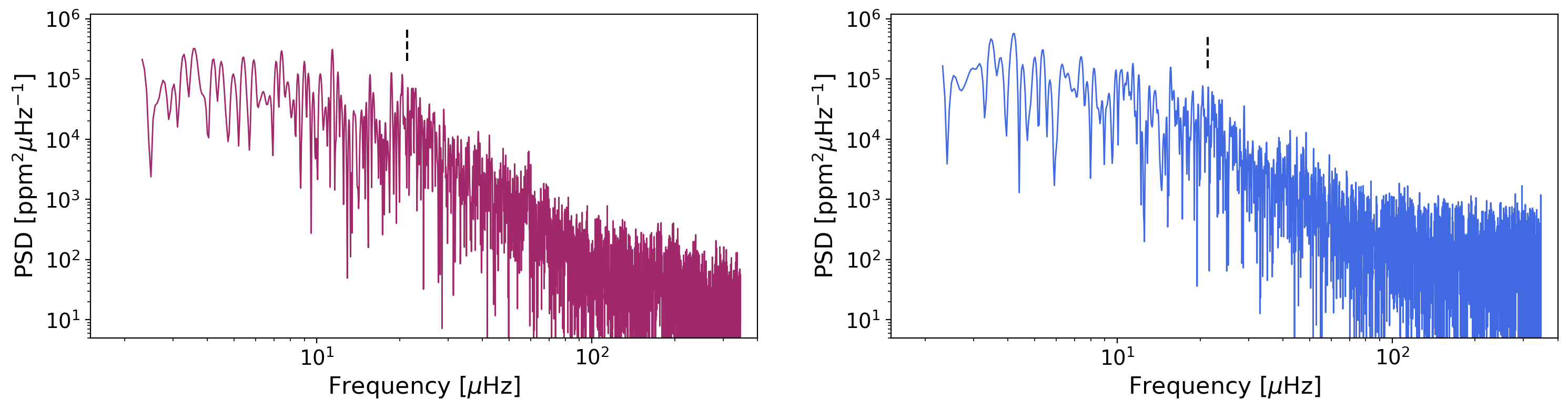}
    \caption{PSDs of the light curves in Fig.~\ref{fig:halo_lc}, with the power spectrum of the custom SAP light curve and the filtered halo light curve on the left and right, respectively. Dashed line marks the $\nu_{\rm max}$ inferred from the custom light curve at $21.27\,\micro$Hz.}
    \label{fig:halo_psd}
\end{figure*}

Halo photometry was applied to most of the brightest stars observed by TESS by \citet{Rudrasingam-2026}, but $\gamma$~Per ($V=2.93$) was below their brightness cut-off of $V=2.66$. We thus extracted the halo photometry of the star and compared it to the PDCSAP and our custom SAP light curves. As Fig.~\ref{fig:halo_lc} shows for the combined Sector 85--86 observations, the halo light curve has a larger scatter than the custom SAP photometry (the same is true for S58) as expected for stars fainter than $V=2.55$ \citep{Rudrasingam-2026}. It also shows increased scatter and segments in S86 with strong systematics similar to what we found in the SAP photometry, which we filtered out. The power spectrum of the halo light curve in Fig.~\ref{fig:halo_psd} is largely similar to that of the SAP photometry, with the power excess identifiable at the same position, confirming our detection.

\section{SONG RV Curves and the Interpolation of the SONG 2017 Data}
\begin{table}[h!]
    \renewcommand{\arraystretch}{1.3}
    \centering
    \caption{Sample table of the SONG 2017 observations. The full table is available in machine-readable format in the electronic version.}
    \label{tab:lc_song}
    \begin{tabular}{l l l}
        \hline 
        BJD--2450000 & RV & RV error\\
        (days) & (m/s) & (m/s) \\
        \hline
        $650.518360$ & $6158.097833$ & $3.9096$ \\
        $650.519820$ & $6158.176333$ & $3.8357$ \\
        $650.521250$ & $6154.589833$ & $3.8923$ \\
        $650.522690$ & $6154.332533$ & $4.0572$ \\
        $650.524130$ & $6151.372833$ & $3.7540$ \\
        $650.525570$ & $6152.578833$ & $3.7901$ \\
        $650.527020$ & $6153.090533$ & $3.7701$ \\
        $650.528450$ & $6149.276933$ & $3.7068$ \\
        $650.529890$ & $6160.100033$ & $3.8636$ \\
        $650.531330$ & $6160.679533$ & $3.8544$ \\
        \multicolumn{3}{l}{...}\\
        \hline
    \end{tabular}
\end{table}

\begin{table}[h!]
    \renewcommand{\arraystretch}{1.3}
    \centering
    \caption{Sample table of the SONG 2024 observations. The full table is available in machine-readable format in the electronic version.}
    \label{tab:lc}
    \begin{tabular}{l l l}
        \hline 
        BJD--2450000 & RV & RV error\\
        (days) & (m/s) & (m/s) \\
        \hline
        $10641.370738$ & $10074.360928$ & $2.8444$ \\
        $10641.372873$ & $10077.528535$ & $3.1425$ \\
        $10641.374996$ & $10072.080766$ & $3.9385$ \\
        $10641.377128$ & $10075.577086$ & $3.6412$ \\
        $10641.379277$ & $10073.231474$ & $3.4827$ \\
        $10641.381420$ & $10075.593677$ & $3.7268$ \\
        $10641.383567$ & $10071.303675$ & $3.1254$ \\
        $10641.385692$ & $10070.389412$ & $4.0114$ \\
        $10641.387829$ & $10071.113131$ & $4.9086$ \\
        $10643.476052$ & $10061.811643$ & $2.3493$ \\
        \multicolumn{3}{l}{...}\\
        \hline
    \end{tabular}
\end{table}

As described in Section~\ref{sec:song_data}, the 2017 SONG RV data was dominated by a beating pattern and strong daily aliasing, which we aimed to resolve by Gaussian Processes interpolation as detailed in Section~\ref{sec:song_gp}.

\begin{figure*}[h!]
    \centering
    \includegraphics[width=\linewidth]{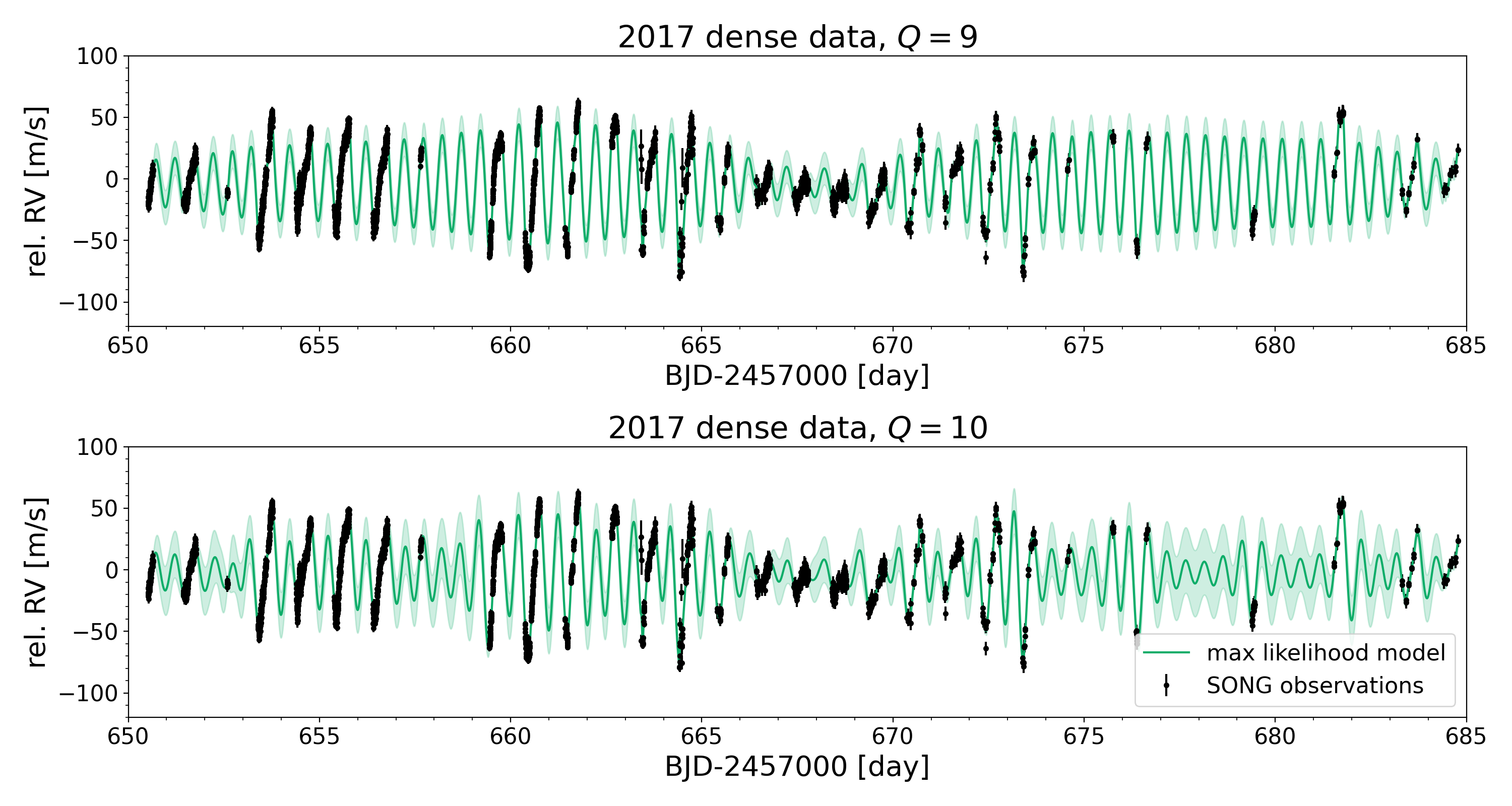}
    \caption{Two solutions for the \texttt{tinygp} fit of 2017 dense SONG observations (black dots). The maximum likelihood model is shown with green lines, while green shaded area refers to the standard deviation of the model.}
    \label{fig:song_2017_dense}
\end{figure*}

\begin{figure*}[h!]
    \centering
    \includegraphics[width=\linewidth]{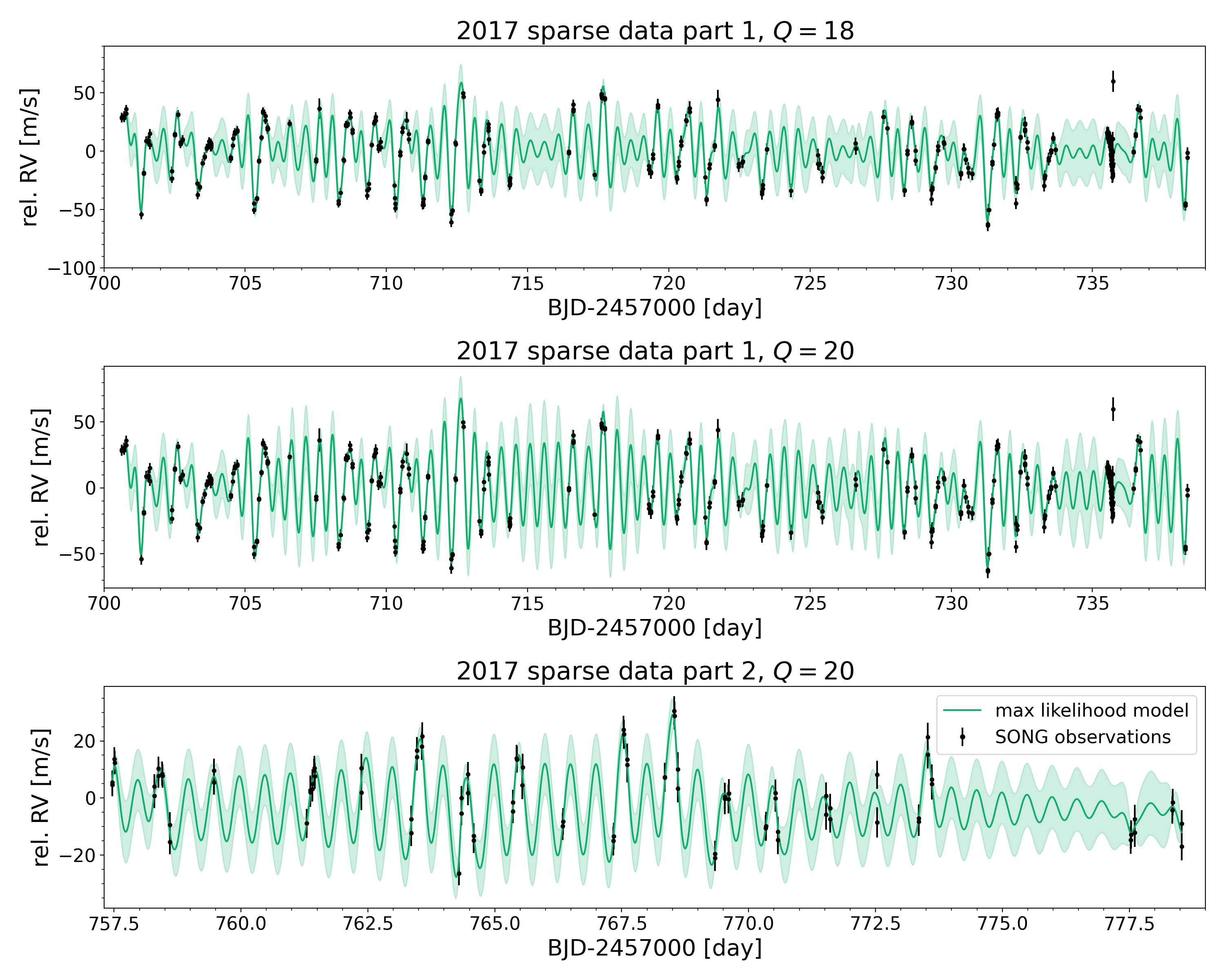}
    \caption{The first two panels present \texttt{tinygp} fits of the first part of the sparse section of the 2017 SONG observations. The bottom plot shows fit of the sparse section part 2 of the 2017 campaign. In all cases the black dots are the SONG observations, the maximum likelihood model is shown with green lines, while green shaded area refers to the standard deviation of the model.}
    \label{fig:song_2017_sparse}
\end{figure*}

We first divided the data into three segments, one dense and two sparse ones as shown in Fig.~\ref{fig:song_2017_all}. Nevertheless, even the dense section was too sparse for a definite \texttt{tinygp} fit, consequently the two most likely solutions are shown in Fig.~\ref{fig:song_2017_dense} (with $Q=9.0$ and $10.0$). Likewise, we provided models for the two sparse sections in Fig.~\ref{fig:song_2017_sparse}, specifically two for the part 1 with $Q=18.0$ and $Q=20$, and one for part 2 with $Q=20.0$.

\begin{figure*}
    \centering
    \includegraphics[width=\linewidth]{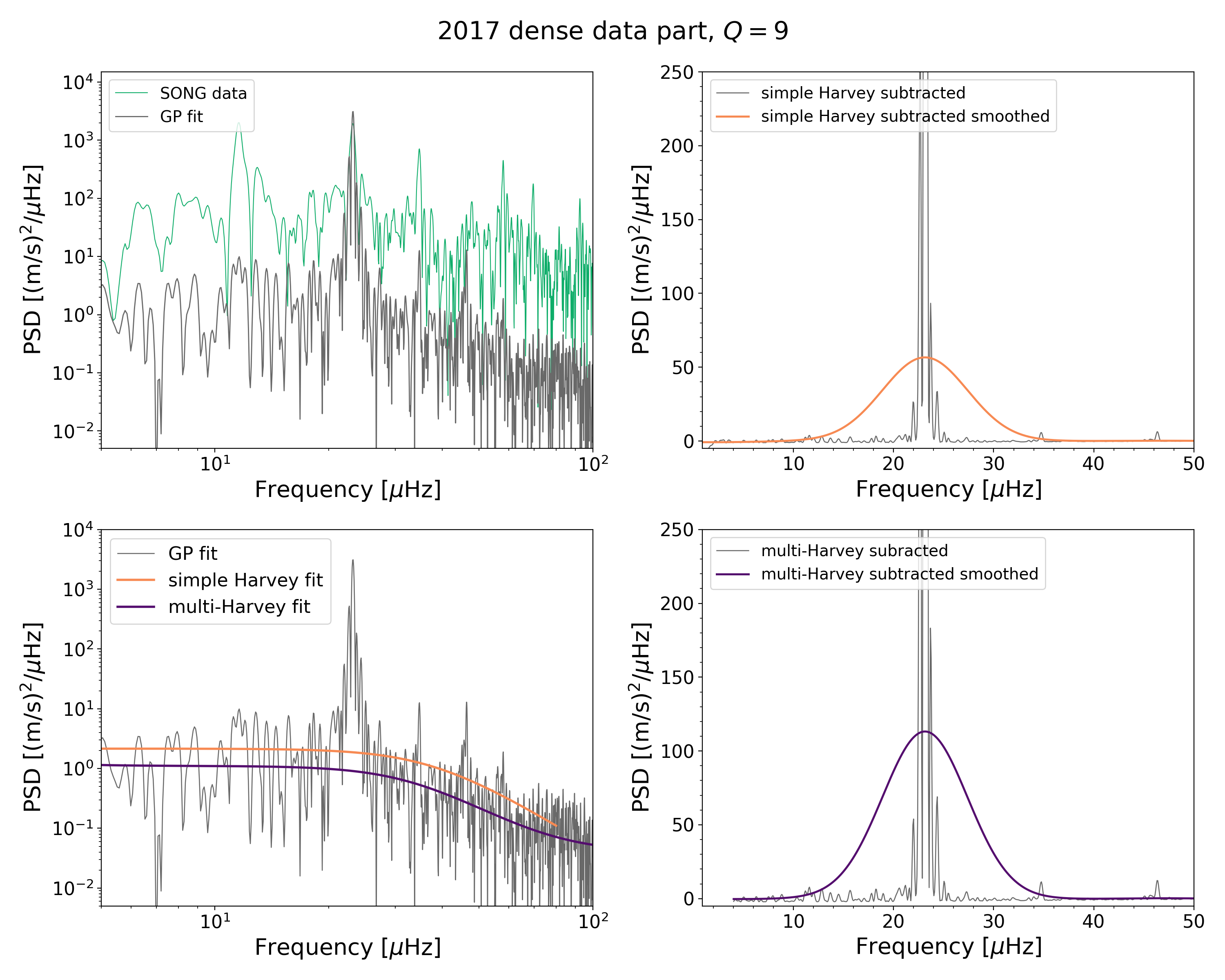}
    \caption{PSDs of first solution for the SONG 2017 dense segment with background fits. Grey lines correspond to the \texttt{tinygp} model in the top plot of Fig.~\ref{fig:song_2017_dense}.}
    \label{fig:song_2017_densea_backfits}
\end{figure*}

\begin{figure*}
    \centering
    \includegraphics[width=\linewidth]{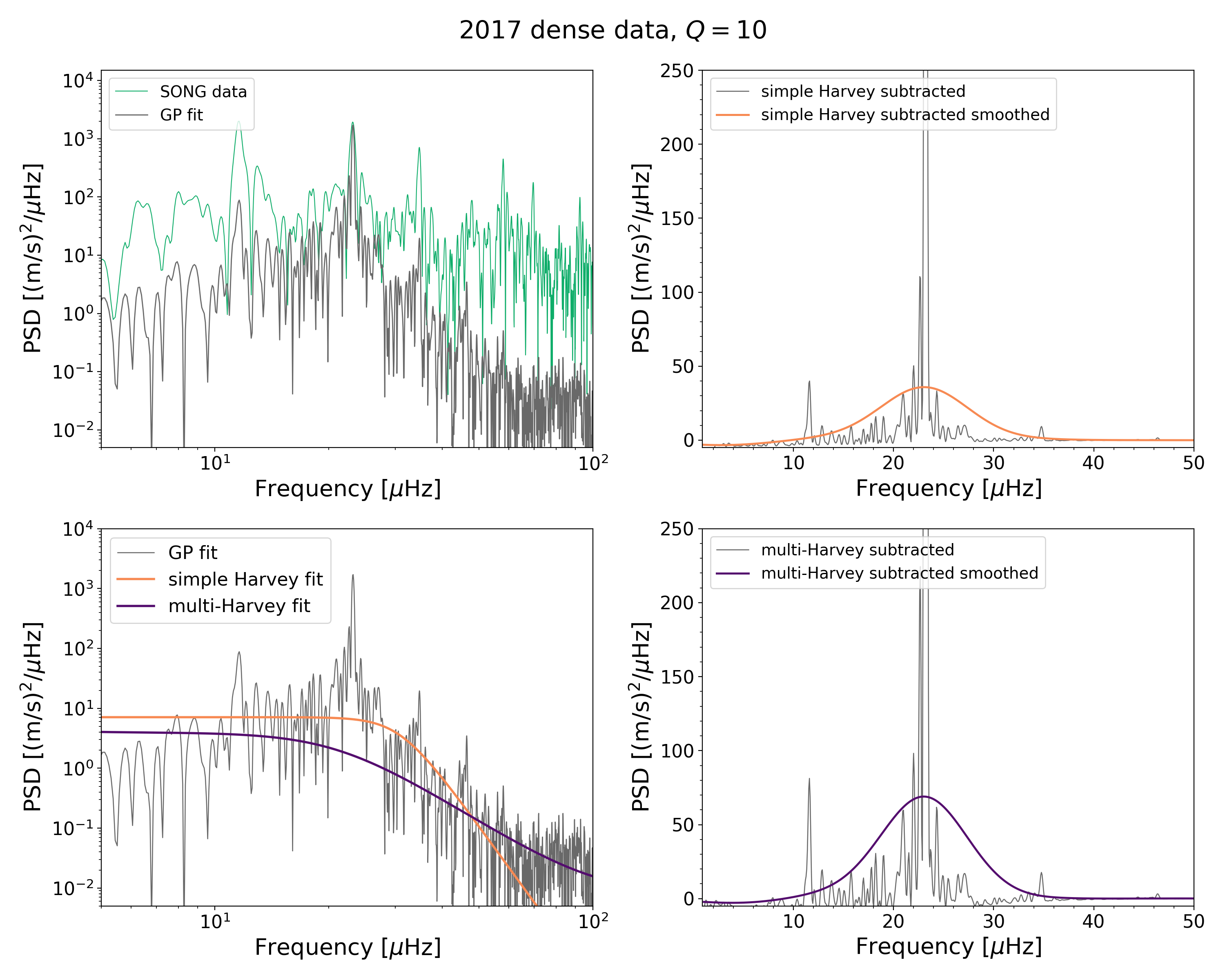}
    \caption{PSDs of second solution for the SONG 2017 dense segment with background fits. Grey lines correspond to the \texttt{tinygp} model in the bottom plot of Fig.~\ref{fig:song_2017_dense}.}
    \label{fig:song_2017_denseb_backfits}
\end{figure*}

\begin{figure*}[h!]
    \centering
    \includegraphics[width=\linewidth]{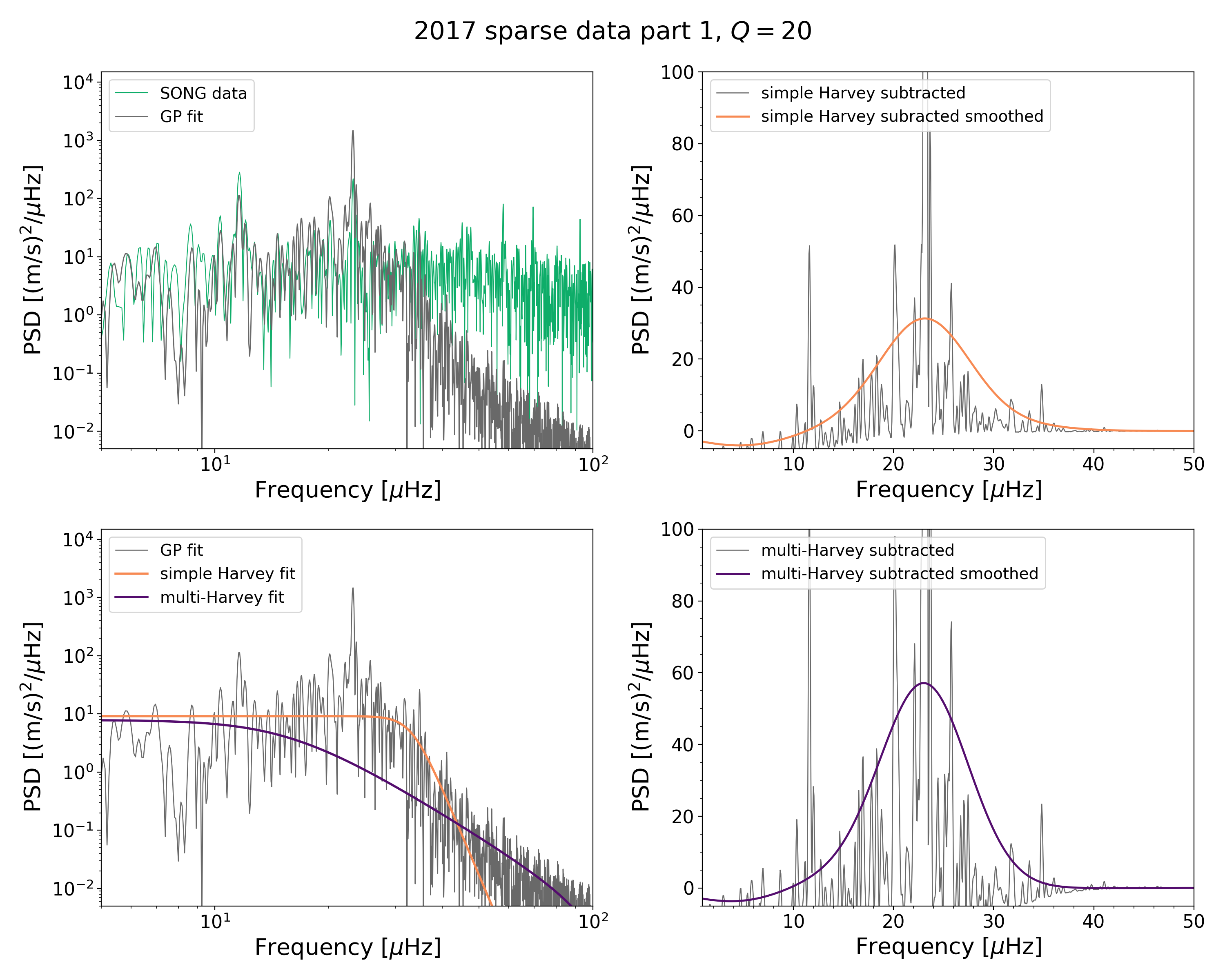}
    \caption{RV amplitude determination of the sparse section part 1 with $Q=18$ of the 2017 SONG observations, the corresponding RV is shown in the top panel of Fig.~\ref{fig:song_2017_sparse}.}
    \label{fig:song_2017_sparse_p1b_backfits}
\end{figure*}

\begin{figure*}
    \includegraphics[width=\linewidth]{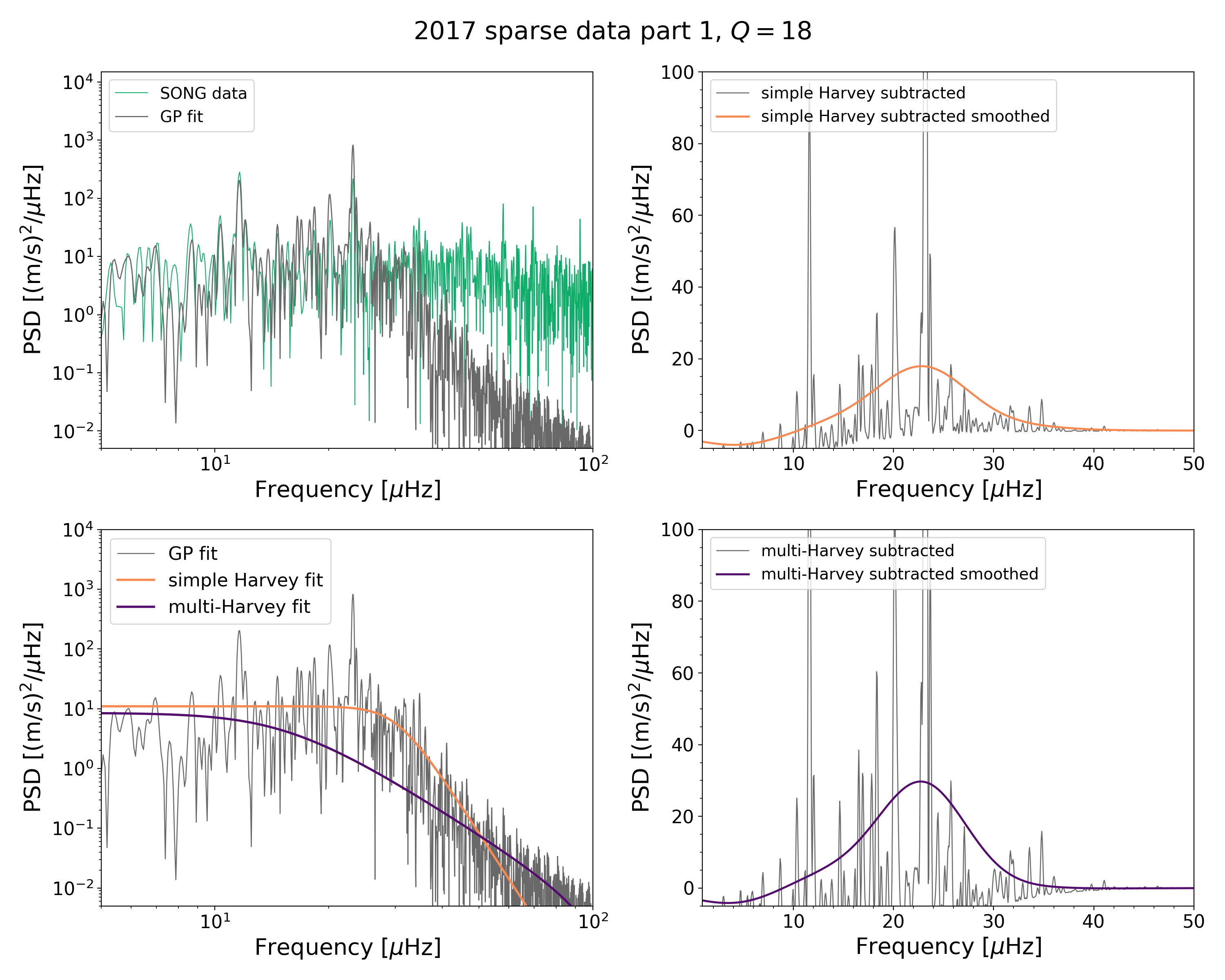}
    \caption{RV amplitude determination of the sparse section part 1 with $Q=20$ of the 2017 SONG observations, the corresponding RV is shown in the middle panel of Fig.~\ref{fig:song_2017_sparse}.}
    \label{fig:song_2017_sparse_p1a_backfits}
\end{figure*}

\begin{figure*}
    \includegraphics[width=\linewidth]{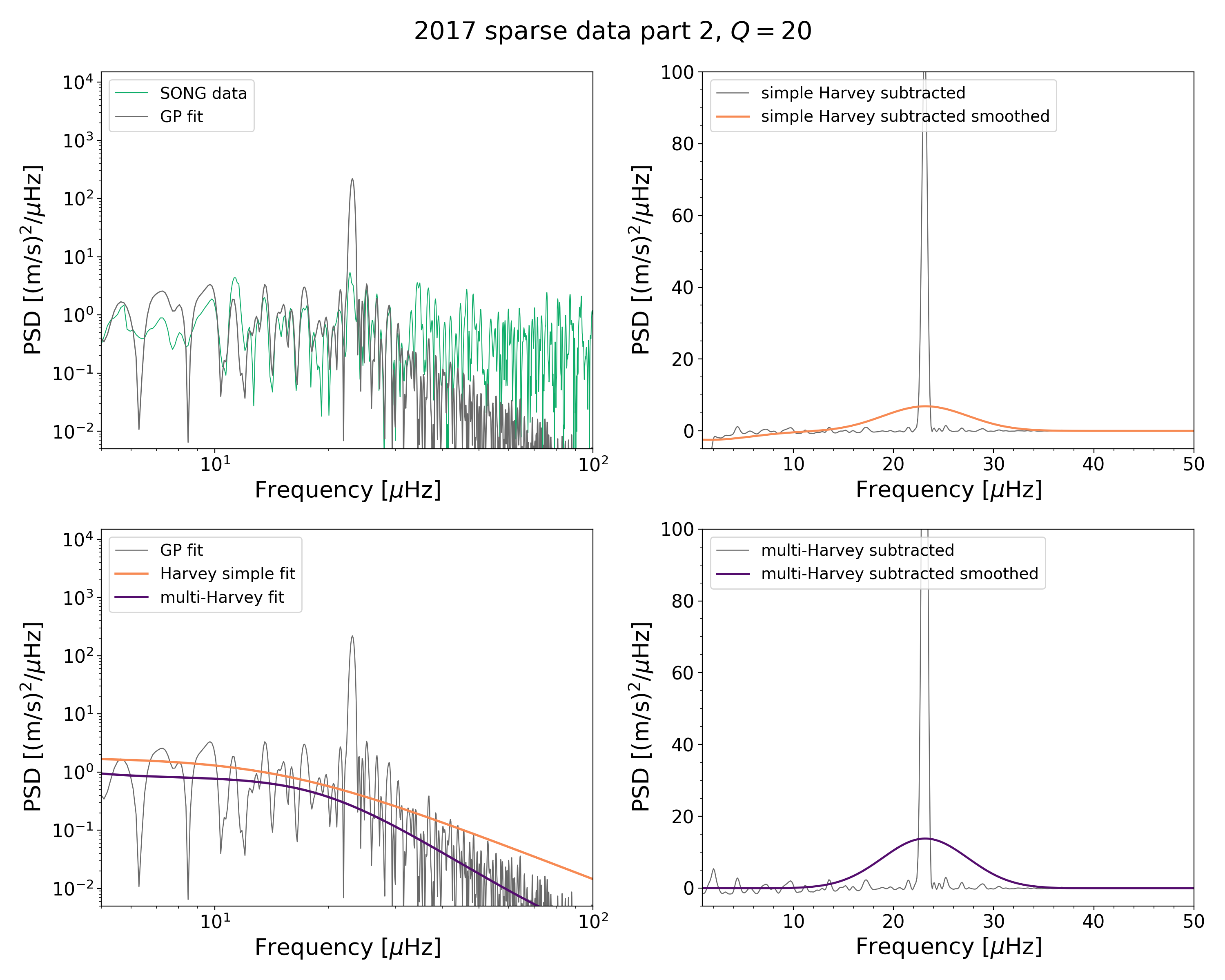}
    \caption{RV amplitude determination of the sparse section part 2 of the 2017 SONG observations, the corresponding RV is shown in the bottom panel of Fig.~\ref{fig:song_2017_sparse}.}
    \label{fig:song_2017_sparse_p2_backfits}
\end{figure*}

Like the middle and bottom panel of Fig.~\ref{fig:song_2024}, the PSDs of the original data and the GP models are presented green and dark grey lines in the left panels, respectively, and the orange lines show the granulation background fit with single Harvey function, the purple ones are ourmulti-component Harvey-like functions fit in Figs.~\ref{fig:song_2017_densea_backfits}, \ref{fig:song_2017_denseb_backfits}, \ref{fig:song_2017_sparse_p1a_backfits}, \ref{fig:song_2017_sparse_p1b_backfits}, and \ref{fig:song_2017_sparse_p2_backfits}. 
On the right panels we present GP model PSDs after subtracting each background model, and the results of the Gaussian smoothing are shown with thick orange and purple lines for each.

\section{Stars Used for Comparison}
To check the validity of our results, we compared the derived parameters to other published ones. Since these values are scattered among several different works, we collected the parameters of these stars in Table~\ref{tab:stars_data} with their respective references, which we hope will help future investigations in the this field.

\begin{table*}[h]
    \centering
    \renewcommand{\arraystretch}{1.3}
    \small{
    \begin{tabular}{l|c|c|c|c}
        Name & $T_{\rm eff} \ [K]$ & Mass [$M_\odot$] & $\nu_{\rm max}$ [$\micro$Hz] & $A_{\rm vel}$ [cm/s] \\ \hline \hline
        
        $\beta$ Aql A & $5155\pm15$ & $1.24\pm0.02$ & $425\pm5$ & $49.7\pm4.4$ \\
        HD188512 & \citet{Soubiran_2022} & \citet{Kjeldsen-2025} & \citet{Kjeldsen-2025} & \citet{Kjeldsen-2025} \\ \hline

        \multirow{2}{*}{$\alpha$ Cen A} & $5804\pm13$ & $1.0788\pm0.0029$ & $2400\pm240$ & $22.5$ \\
        & \citet{Soubiran_2022} & \citet{Akeson_2021} & \citet{Kjeldsen_2008} & \citet{Kjeldsen_2008} \\ \hline
        
        \multirow{2}{*}{$\alpha$ Cen B} & $5207\pm12$ & $0.9092\pm0.0025$ & $4100\pm410$ & $7.2$ \\
        & \citet{Soubiran_2022} & \citet{Akeson_2021} & \citet{Kjeldsen_2008} & \citet{Kjeldsen_2008} \\ \hline

        $\tau$ Cet & $5339\pm19$ & $0.783\pm0.012$ & $4490\pm100$ & $11.2\pm0.8$ \\
        HD10700 & \citet{Hojjatpanah_2019} & \citet{Teixeira-2009} & \citet{Teixeira-2009} & \citet{Teixeira-2009} \\ \hline

        $\alpha$ For A & $6240$ & $1.33\pm0.01$ & $1100\pm110$ & $34.8\pm0.9$ \\
        HD20010A & \citet{Santos_2001} & \citet{Santos_2001} & \citet{Kjeldsen_2008} & \citet{Kjeldsen_2008} \\ \hline
        
        $\beta$ Hyi & $5917\pm25$ & $1.127\pm0.054$ & $1000\pm100$ & $41.9\pm0.9$ \\
        HD2151 & \citet{Soubiran_2022} & \citet{Metcalfe_2024} & \citet{Bedding-2007} & \citet{Bedding-2007} \\ \hline

	    $\epsilon$ Ind A & $4700\pm65$ & $0.782\pm0.023$ & $5305\pm176$ & $2.6\pm0.5$ \\
        HD209100 & \citet{Lundkvist_2024} & \citet{Lundkvist_2024} & \citet{Campante-2024} & \citet{Campante-2024} \\ \hline
        
        $\nu$ Ind & $5318\pm80$ & $0.85$ & $320\pm32$ & $64.7\pm0.9$ \\
        HD211998 & \citet{Fuhrmann_2021} & \citet{Fuhrmann_2021} & \citet{Bedding-2006} & \citet{Kjeldsen_2008} \\ \hline
        
        $70$ Oph A & $5370.31\pm1$ & $0.90\pm0.10$ & $4500\pm500$ & $13.9$ \\
        HD165341 & \citet{Picotti_2020} & \citet{Picotti_2020} &  \citet{Carrier-2006} & \citet{Carrier-2006} \\ \hline

        $\delta$ Pav & $5609\pm8$ & $1.07\pm0.01$ & $2300\pm300$ & $19.5\pm0.4$ \\
        HD190248 & \citet{CarvalhoSilva_2025} & \citet{CarvalhoSilva_2025} & \citet{Kjeldsen_2008} & \citet{Kjeldsen_2008} \\ \hline
        
        $18$ Sco & $5824\pm30$ & $1.01\pm 0.02$ & $3000\pm200$ & $20$ \\
        HD146233 & \citet{Soubiran_2024} & \citet{Karovicova_2022} & \citet{Bazot-2011} & \citet{Bazot-2011} \\ \hline
        
        $\gamma$ Ser & $6296\pm16$ & $1.21$ & $1600\pm160$ & $34.2\pm1.3$ \\
        HD142860 & \citet{Soubiran_2022} & \citet{AguileraGomez_2018} & \citet{Kjeldsen_2008} & \citet{Kjeldsen_2008} \\ \hline

        $\epsilon$ Tau & $4950\pm22$ & $2.458\pm0.073$ & $56.4\pm1.1$ & $94\pm4$ \\
        HD28305 & \citet{Arentoft_2019} & \citet{Arentoft_2019} & \citet{Arentoft_2019} & \citet{Arentoft_2019} \\ \hline

        \multirow{2}{*}{HD35833} & $5684\pm65$ & $1.42\pm0.04$ & $596\pm17$ & $111\pm9$ \\
        & \citet{Fouesneau_2023} 
        & \citet{Gupta_2022} & \citet{Gupta_2022} & \citet{Gupta_2022} \\ \hline

        \multirow{2}{*}{HD219134} & $4817\pm62$ & $0.763\pm0.034$ & $4651\pm301$ & $4.23\pm0.41$ \\
        & \citet{Rosenthal-2021} 
        & \citet{Li-2025} & \citet{Li-2025} & \citet{Li-2025} \\ \hline

        Procyon A & $6478.25\pm35.20$ & $1.478 \pm 0.012$ & $1014\pm9$ & $38.1 \pm 1.3$ \\
        HD 61421 & \citet{Nepal_2023} & \citet{Bond_2015} & \citet{Huber_2011_2} & \citet{Arentoft_2008} \\ \hline

        \multirow{2}{*}{Sun} & $5772$ & \multirow{2}{*}{$1$} & $3100\pm40$ & $16.0\pm0.4$ \\
        & \citet{Prsa-B3-2016} & & \citet{Hekker_2020} & \citet{Kjeldsen-2025} \\ \hline
    \end{tabular}
    } \vspace{0.1cm}
    \caption{The astrophysical parameters of other stars we used for comparison in Figs.~\ref{fig:comp_teff} and \ref{fig:comp_numax} with their respective references. Additionally to their names, we indicated their Henry Draper Catalogue ID, if applicable.}
    \label{tab:stars_data}
\end{table*}

\end{appendix}

\end{document}